\documentclass[a4paper,11pt]{article}
\usepackage{jheppub} 
\usepackage{amsmath, amssymb, amsthm, epsfig, fancyhdr,epsfig,multirow}
\usepackage[utf8]{inputenc}
\usepackage{amsmath}
\usepackage{amsfonts}
\usepackage{amssymb}
\usepackage{xcolor}
\usepackage{comment}
\usepackage{subcaption}
\usepackage[normalem]{ulem}
\usepackage{tabularx}
\usepackage{comment}
\usepackage{float}
\usepackage{graphicx}
\usepackage{appendix}
\usepackage[most]{tcolorbox}
\usepackage{physics}
\usepackage{mathtools}

\newcommand\Mycomb[2][^n]{\prescript{#1\mkern-0.5mu}{}C_{#2}}

\title{Singularities in Cosmological Loop Correlators}

\author[1]{Supritha Bhowmick,}
\author[2]{Mang Hei Gordon Lee,}
\author[1]{Diptimoy Ghosh,}
\author[1]{Farman Ullah}
\affiliation[1]{Indian Institute of Science Education and Research Pune, Pune 411008, India}
\affiliation[2]{Leung Center for Cosmology and Particle Astrophysics, National Taiwan University,
Taipei 10617, Taiwan.}

\emailAdd{supritha.bhowmick@students.iiserpune.ac.in} 
\emailAdd{mhglee@ntu.edu.tw}
\emailAdd{diptimoy.ghosh@gmail.com} 
\emailAdd{farman.ullah@students.iiserpune.ac.in}

\abstract{In this work we perform a systematic study of the singularity structure of inflationary correlations at 1-loop. We explicitly compute a few diagrams and find a pattern emerging in the singularities produced. Motivated by this, we derive diagrammatic rules to extract the singularities of any two-site 1-loop diagram. Using these rules, the poles and branch cuts produced can be predicted by simply identifying the energies flowing through certain subgraphs, without having to perform complicated integrals. We demonstrate how these rules follow by analyzing the general structure of the time and momentum integrals of the correlators. An interesting feature of de-Sitter correlators at 1-loop is the presence of an \textit{off-shell} total energy branch point, which is present in dimensional regularization as well as cutoff regularization. We probe the source of this branch cut in detail, while revisiting the cosmological KLN theorem \cite{AguiSalcedo:2023nds} in this context. Finally, we show that the branch cuts produced in a renormalised correlator always repackage themselves in a dilatation invariant form to produce logarithms of ratios of comoving scales.}

\begin{document}
\maketitle
\flushbottom

\section{Introduction}

About 20 years ago, Weinberg computed the one loop correction to the primordial power spectrum and found the following \cite{Weinberg:2005vy}:
\begin{equation}
    \langle\phi_{\vec{k}}\phi_{\vec{k'}}\rangle^{\text{1-loop}}\sim\delta^{(3)}(\vec{k}+\vec{k'})\frac{1}{k^3}\log(k/\mu).\label{eqn:logk}
\end{equation}
If we accept this result as true, this has huge and worrying implications on the physics of inflation, the leading paradigm for generating primordial fluctuations. First, the choice of a suitable renormalization scale requires the numerator of the argument of the logarithm to be a physical scale. Rewriting it as $\log\left\{k_{\text{phys}} ~a/\mu \right\}$, we see that there is an explicit dependence of the power spectrum on the scale factor. This is a serious problem : any physical observable should not depend on the scale factor, since it simply corresponds to a rescaling of coordinates. Secondly, for long wavelength modes, i.e. $k\rightarrow 0$, the loop corrections to the power spectrum could become unsuppressed, in which case we can no longer use perturbation theory as a predictive tool for inflation \cite{Senatore:2009cf}.

 Fortunately this is not the correct answer. In \cite{Senatore:2009cf} it was found that the obtained logarithm was a mistake due to an improper implementation of the dimensional regularization scheme, which led to an answer with unphysical predictions. After implementing dimensional regularization in de Sitter (dS) correctly, by using $d(=3+\delta)$ dimensional mode functions, the answer turns out to be:
\begin{equation}
    \langle\phi_{\vec{k}}\phi_{\vec{k'}}\rangle^{\text{1-loop}}\sim\delta^{(3)}(\vec{k}+\vec{k'})\frac{1}{k^3}\log(H/\mu).
    \label{eqn:logH}
\end{equation}
The worrying features from Weinberg's calculation are no longer present. 

There is an important lesson for us in this story: properly understanding loop effects is essential for understanding the predictions as well as the robustness of inflationary theories. In many cases secular growth are a genuine feature of the theory and has to be properly resummed (for instance, through stochastic inflation)\cite{Ford:1984hs,Antoniadis:1985pj,Starobinsky:1994bd,Dolgov:2005se,Marolf:2010zp,Burgess:2010dd, Tsamis:2005hd, Prokopec:2007ak,Woodard:2025cez,Woodard:2025smz,Foraci:2024cwi,Foraci:2024vng,Miao:2024shs}; loop effects can also appreciably affect predictions on non-Gaussianities \cite{Riotto:2008mv,Assassi:2012et,Gorbenko:2019rza,Green:2020txs,Cohen:2021fzf,Wang:2021qez,Lee:2023jby}, or lead to interesting predictions such as production of primordial black holes \cite{Sasaki:2018dmp,Kristiano:2022maq,Riotto:2023hoz,Kristiano:2023scm,Riotto:2023gpm,Choudhury:2023jlt,Franciolini:2023agm}.

Aside from secular divergences, there are many other reasons to care about loop effects in dS and inflation. One of the important lessons from the S-matrix community is that physical constraints are often manifest as analyticity properties in the observable: unitarity, locality and causality all leave imprints in the complex plane of the kinematics of the amplitude \cite{Eden:1966dnq, Martin:1969ina,Elvang:2015rqa,Nussenzveig:1972tcd,Sashanotes, Mizera:2023tfe}. Since loop effects may introduce branch cuts (through transcendental functions such as polylogarithms), understanding loop corrections can lead us to a better understanding of the analytic structure of inflationary correlators, and may bring us a step closer towards bootstrapping cosmological observables. 

While in-in correlators are the observable of choice for many cosmologists, the analytic properties of the wavefunction is better understood~\cite{Benincasa:2018ssx,Benincasa:2021qcb,Salcedo:2022aal,Benincasa:2024ptf}. In cases where IR divergences are absent (for instance the effective field theory of inflation~\cite{Cheung:2007st}), the wavefunction or correlator calculations in dS or inflation can be related to those in flat space by taking derivatives~\cite{Hillman:2021bnk,Lee:2023jby}. We know that the singularities of the flat space wavefunction are given by the \textit{energy conservation condition}: poles and branch cuts occur when energies entering a subgraph of a Feynman diagram vanishes \cite{Salcedo:2022aal}. In addition, in-in correlators can be computed from wavefunction coefficients, so their singularities should be related to each other. This give us an easy way to study the analyticity properties of in-in correlators in some phenomenologically relevant inflationary theories. An effort in this direction was made in \cite{AguiSalcedo:2023nds}. 

In this paper we will re-examine the analytic structure of in-in correlators carefully, by studying the features of the time integrals for the in-in correlators and regularizing the integrals properly. This paper is organized as follow: we compute a few examples of correlation functions  at 1-loop with 2 sites in Sec.~\ref{Sec:Explicit_Comp}. Motivated by the trend in the results, we deduce diagrammatic rules to extract the singularity structure of loop corrections in Sec.~\ref{Sec:Singularity_structure}. These rules may be used to bypass explicit computations and obtain all possible poles and branch cuts simply based on the energy injection into the loop sites and various subgraphs. This is similar to how Feynman rules can be followed to write down integral expressions for a given diagram without evaluating the full in-in expression. In Sec.~\ref{Sec:Working Principle} we consider the structure of a general $n$-point function at 1-loop with 2-sites and study the expected log structure arising once the loop integrals are performed, demonstrating how the previously mentioned diagrammatic rules follow. In Sec.~\ref{Sec:Renormalization} we revisit the cosmological KLN theorem \cite{AguiSalcedo:2023nds}, which forbids the presence of total energy branch points in off-shell cosmological correlators of massless fields. We discuss a loophole which allow us to bypass the cosmological KLN theorem: namely, de Sitter regularization schemes (including dimensional regularization with $d$-dimensional mode function as well as time dependent cutoff regularization) can give rise to terms which generates total energy branch points. We also show that all branch cuts produced will always repackage themselves in a renormalised correlator to produce logarithms of ratios of comoving scales, thereby not producing any scale-breaking terms.\\

The main results of this work are as follows,
 \begin{itemize}
     \item The singularity structure of any \textit{off-shell} correlation function at 1-loop with 2-sites can be extracted without explicit computation, using diagrammatic rules (see Sec.~\ref{Sec:Singularity_structure}). To proceed, one must identify certain subgraphs in the diagram, which we call the \textit{``loop"-}, the \textit{``left"-} and the \textit{``right"-} subgraphs, defined in Sec.~\ref{Sec:Singularity_structure}.
     
     \begin{itemize}
         \item If the energy flowing into the \textit{loop}-subgraph through the left and right loop site is $\mathcal{S}_L$ and $\mathcal{S}_R$ respectively, then the loop correction will feature a logarithm of $\mathcal{S}_L+s$ and $\mathcal{S}_R+s$.
         \item If the energy flowing through the \textit{left-} (\textit{right-}) subgraph through all vertices except the right (left) loop site is $\mathcal{S}_{L_i}$ ($\mathcal{S}_{R_j}$), then the loop correction will feature a logarithm of $\mathcal{S}_{L_i}+s$ ($\mathcal{S}_{R_j}+s$).
     \end{itemize}
     These results hold for correlation functions in flat space as well as de-Sitter. In de Sitter, for the case involving only massless scalars, the rules hold for interactions with derivative(s) acting on each field insertion. For correlators in flat space and de Sitter correlators where the fields running in the loop are conformally coupled scalars, our rules work with polynomial as well as derivative interactions.
     
     \item Additionally, correlation functions in de-Sitter feature logarithm of total energy from time integrals due to $\mathcal{O}(\delta)$ contributions from modes and measure. 

     \item In a renormalised correlator, all branch cuts will combine to produce logarithms of ratios of comoving scales, such that final answer is scale invariant.
     
 \end{itemize}

\section{Notations and Conventions}
\label{Sec:Notations_Conventions}
The spatially flat FRW metric is given as,
\begin{align}
    ds^{2}=a^{2}(-d\tau^{2}+d\vec{x}^{2}) \,,
\end{align}
where $a$ is the scale factor and $\tau$ is conformal time. For a flat de Sitter FRW background ($a=-1/(H\tau)$) we get,
\begin{align}
ds^{2}=\frac{-d\tau^{2}+d\vec{x}^{2}}{\left(H\tau\right)^{2}}   \,, 
\end{align}
where $H$ is the Hubble parameter. The FRW time coordinate is denoted by $t$. Derivatives with respect to conformal time are denoted by a prime. The external momenta are denoted by $\{\vec{k}_i\}$ and exchange momenta by $\{\vec{p}_i\}$ with $\vec{p}_1$ \& $\vec{p}_2$ reserved to label the loop arms. For \textit{on-shell} propagators, the modulus of momenta are referred to as energies and denoted by momentum label without the vector sign e.g., $k_i=|\vec{k}_i|$. However, we will often treat the \textit{energies} of external legs as independent variables (this is the \textit{off-shell} limit of the correlator) and in this case the energies are denoted by $\{\omega_i\}$. The energy entering the loop site is denoted by $s$. Often, we will denote sum of off-shell energies in the following manner : $\omega_1+\omega_2+\omega_5=\omega_{125}$ and so on. 

The correlation functions are computed via the in-in formalism. To be precise, we employ in-in Feynman rules to compute the diagrams. In this work, the notation we follow for the diagrams are as follows : vertices drawn at $\tau<0$ come from the time ordered part of the in-in correlator, and vertices drawn at $\tau>0$ come from anti-time ordered part. These vertices are also called ``right" (or `$+$') and ``left" (or `$-$') vertices respectively (\cite{Weinberg:2005vy, Chen:2017ryl}). Of course a $n$-point correlator will have contributions from diagrams with all possible combinations of left and right vertices. However not all of these diagram need to be computed since half of these combinations are obtained from complex conjugation of the other half.

Given a mode function $f_k(\tau)$, the propagators are defined as follow: 
\begin{align}
    G_{++}(k;\tau_1,\tau_2)&=f_k^\ast(\tau_1)f_k(\tau_2)\Theta(\tau_1-\tau_2)+f_k(\tau_1)f_k^\ast(\tau_2)\Theta(\tau_2-\tau_1),\\
    G_{+-}(k;\tau_1,\tau_2)&=f_k(\tau_1)f^\ast_k(\tau_2),\\
    G_{--}(k;\tau_1,\tau_2)&=G_{++}^\ast(k;\tau_1,\tau_2),\\
     G_{-+}(k;\tau_1,\tau_2)&=G_{+-}^\ast(k;\tau_1,\tau_2), \label{Eqn: Propagators}
\end{align}
where $+$ and $-$ in the subscripts denotes whether $\tau_1,\tau_2$ are greater (or smaller) than $0$.

\paragraph{Dimensional regularization}
Throughout this paper we will employ dimensional regularization (dim reg) to tame the divergences in loop integrals. In flat space this amounts to computing the loop integral in $d$-dimensions, then setting $d=3+\delta$. In de Sitter this is more subtle: mode functions in dS depend on the dimension. Therefore, when we compute correlators with dimensional regularization, we need to use the appropriate $d$-dimensional mode functions as well, as they may include crucial contributions to the structure of the correlator. A famous example would be the $\log(k)$ term mentioned in the introduction: using the incorrect mode function would lead us to conclude that there is a secular divergences in the one loop correction of the power spectrum \cite{Weinberg:2005vy}, and this is tamed by properly regulating the mode functions as well \cite{Senatore:2009cf}.

For correlators in flat space, since there is no dimensional dependence of modes, simple $3$-dimensional flat space modes, given by,
\begin{align}
    f_{k}(\tau)=\frac{1}{\sqrt{2k}} \exp(i k \tau) \,,
\end{align}
are employed in loop computations.

In de Sitter, the mode function in $d$-dimensions for a scalar field with mass $m$ is given by:
\begin{equation}
    f_k(\tau)=(-H\tau)^{d/2}H^{(2)}_{i\nu}(-k\tau),
\end{equation}
where $\nu=\sqrt{\frac{m^2}{H^2}-\frac{d^2}{4}}$. However it is difficult to compute correlators in arbitrary $d$-dimension: in general we would need to compute time integrals and loop momentum integrals with Hankel functions, and these don't usually give nice analytic expressions.

Fortunately for our purposes there are two major simplifications. The first is that we are mainly interested in massless fields ($m^2=0$) or conformally coupled fields ($m^2=2H^2$) in $d=3$. For $d=3$ these Hankel functions reduces to simply plane waves multiplying some polynomials. The second simplification comes from modifying the mass slightly when we do dimensional regularization: on top of setting $d=3+\delta$, we also modify the mass as $m^2\rightarrow m^2+H^2 \left( \frac{3}{2}\delta+\frac{\delta^2}{4} \right)$\cite{Melville:2021lst,Lee:2023jby}\footnote{This is different from the scheme used in \cite{Senatore:2009cf} where the mass is not modified. While the scheme of modifying the mass simplifies the calculations further, one should be careful when employing this modified scheme on gauge fields or gravitons, since introducing a mass may break gauge invariance.}. Under this scheme the mode function simplifies into the following:
\begin{align}
    &f_k(\tau)=(-H\tau)^{\delta/2}\frac{H}{\sqrt{2k^3}}(1-ik\tau)e^{ik\tau}\hspace{17pt}\text{(for massless scalar)} \label{Eqn:massless_dS} \,,\\
    &f_k(\tau)=(-H\tau)^{\delta/2}\frac{\tau}{\sqrt{2k}}e^{ik\tau}\hspace{65pt}\text{(for conformally coupled scalar)} \label{Eqn:conf_coup_dS}\,. 
\end{align}

In practice we would frequently need to expand in $\delta$ to obtain analytic expressions, and this generally produces logarithmic terms. For example, for the massless scalar we have:
\begin{equation}
     f_k(\tau)=\bigg(1+\frac{\delta}{2}\log(-H\tau)\bigg)\frac{H}{\sqrt{2k^3}}(1-ik\tau)e^{ik\tau}+\mathcal{O}(\delta^2).\label{Eqn:Massless_d_dimension}
\end{equation}
Since we are mainly concerned with one loop corrections, it suffices to expand the mode function to $\mathcal{O}(\delta)$, but in principle one would need the higher order terms for higher loop corrections as well.

Similarly, the time integral measure is also dependent on the dimension $d$, and receives correction in dimensional regularization,
\begin{align}
    a^\delta (\tau)=(-H\tau)^{d+\delta}=(-H\tau)^d(1-\delta \log(-H \tau) +\mathcal{O}(\delta^2))\,.
    \label{measure_d}
\end{align}

\paragraph{Off-shell correlators}
In flat space the mode functions are simply plane waves, and as a result the correlators always depends on the total energy entering each vertex (or subgraphs of these vertices), regardless of the number of external legs attached to it. For example, if in a contact diagram there are $N_a$ external legs attached to a vertex $A$, then the correlator only depends on the external energy $|\textbf{k}_i|$ in the combination:
\begin{equation}
    \Omega_A=\sum_{i=1}^{N_a}\sqrt{|\mathbf{k}_i|^2+m_i^2}.
\end{equation}

For de Sitter, if we look at fields with general mass, the correlator will be some complicated function of the energies of individual external legs $k_i=|\mathbf{k}_i|$, rather than depending on simply their sum. However, since we are mainly working with massless and conformally coupled scalars, the mode functions are plane waves multiplied by polynomials of conformal time and energies. This implies that the time integral at each vertex takes the following form:
\begin{equation}
    \int_{-\infty}^{\tau'}d\tau\,\text{Poly}(k,\tau)e^{i\Omega\tau},
\end{equation}
where $\Omega=\sum_{i}k_i$ is the energy entering a vertex if $\tau$ is the first vertex to be integrated over, and the energy entering a subgraph otherwise. $\text{Poly}(k,\tau)$ is simply a polynomial in energies and conformal time. It is not hard to see that after time integration, the denominators will depend only on the external energies in the combination $\Omega$. 

It is useful to define the \textit{off-shell correlator} as an analytic extension of the physical correlator, where the energy of the internal legs remains \textit{on-shell}, i.e. $p_i=\vec{p}_i$, but the energy of the external legs $\omega_i$ is treated as independent from the momentum $\mathbf{k}_i$. Physical answers are obtained when we put the external energy \textit{on-shell}, i.e. take $\omega_i=|\vec{k}_i|$.

Analytic continuation of the off-shell energy has been fruitful in understanding the structure of wavefunctions and correlators \cite{Salcedo:2022aal,AguiSalcedo:2023nds,Lee:2023kno}. In particular, it is possible to keep track of the analytic structure in terms of $\omega$, and in the case of the wavefunction this has led us to rules which determine the possible poles and branch points. It will be helpful for us to work with these off-shell energies here as well. As a result we will work with the off-shell external energies $\omega$ in our dS examples, since it contains interesting features about the analyticity of the correlators.

\section{Explicit Computations}
\label{Sec:Explicit_Comp}

In an effort to better understand the analytic structure of correlators at the loop level, we perform some explicit computations below, in flat space and de-Sitter. We will use dimensional regularisation to regulate loop divergences. Later on we discuss that our results are not just artifacts of the regularisation scheme, and hold true even in cutoff regularisation (See Sec.~\ref{Sec:Loophole}). In this section, we only compute correlator contributions with vertices on the same side of $\tau=0$. Later on in Sec.~\ref{Sec:Unord}, we consider all possible contributions where vertices can be $\tau>0$ or $\tau<0$, and show that the results of our work follow through.

For flat space computations, we will consider a theory with polynomial quartic ($g_4 \phi^4$) and cubic ($g_3 \phi^3$) interactions of a massless scalar $\phi$. For correlators in de Sitter, we will work with interactions involving time derivatives acting on massless scalars and polynomials of conformal scalars ($\sigma$), such as $\dot{\phi}^3$, $\dot{\phi} \sigma^2$ and so on. Also, we will choose to drop overall factors of normalization ($\prod_{ij} \omega_i p_j$) coming from modes as well as numerical constants when stating the results of computations.

\subsection{Correlators in Flat Space}
\label{Sec:Flat Space_examples}

\begin{figure}[h]
\centering
\includegraphics[scale=0.45]{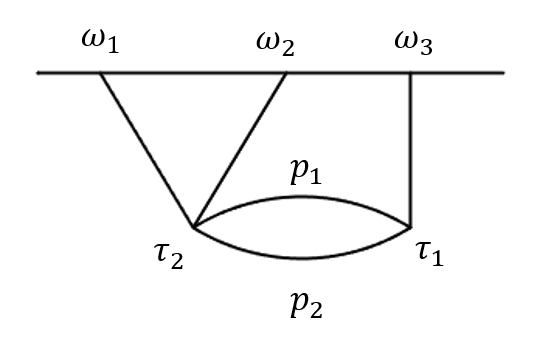}
\caption{Bispectrum at 1-loop with interactions $g_3 \phi^3$ and $g_4\phi^4$.}
\label{Fig:Diagram_1}
\end{figure}

\textbf{Bispectrum at 1-loop with cubic and quartic interaction:} Consider the ($++$) contribution to the bispectrum in the diagram of Fig.~\ref{Fig:Diagram_1}.\\
Let us denote the magnitude of the momentum exchanged between the two sites as $s=|\vec{p}_1+\vec{p}_2|$, and the energy flowing through the vertices as follows,
\begin{align}
    E_L=\omega_1+\omega_2+p_1+p_2 , \hspace{15pt} E_R=\omega_3+p_1+p_2\,.
\end{align}
\\
Evaluation of the time integrals leads to partial and total energy poles,
\begin{align}
     & \langle \phi_{\vec{k}_1} \phi_{\vec{k}_2} \phi_{\vec{k}_3} \rangle^{1-\text{loop}} \sim \frac{g_3 g_4}{\omega_T} \int d^d \vec{p}_1 d^d \vec{p}_2 ~\frac{1}{p_1 p_2}  \frac{1}{ E_L E_R} ( \omega_T+2 p_+)\,.
 \end{align}
\\
where $\omega_T=\omega_1+\omega_2+\omega_3$ and $p_1+p_2=p_+$. Both the partial energies $E_L, E_R$ depend on loop momenta, and these simple poles turn into branch cuts once the loop integrals are carried out, resulting in the following, 
\begin{align}
    & \langle \phi_{\vec{k}_1} \phi_{\vec{k}_2} \phi_{\vec{k}_3} \rangle^{1-\text{loop}} \sim \frac{g_3 g_4}{\omega_T}  \left( \frac{1}{\delta} +\frac{1}{2} \log \left(\frac{\omega_3 \omega_T}{\mu ^2}\right) \right) +\text{NLf}\,, \label{Eqn:Diagram_1}
\end{align}
\\
where $\mu$ is the renormalization scale and NLf denotes Non-Logarithmic finite terms.\\

\begin{figure}[h]
\centering
\includegraphics[scale=0.5]{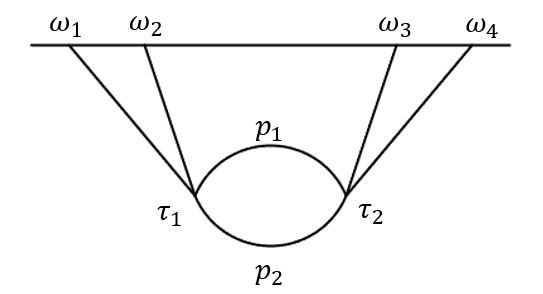}
\caption{Trispectrum at 1-loop with interaction $g_4 \phi^4$.}
\label{Fig:Diagram_2}
\end{figure}

\noindent \textbf{Trispectrum at 1-loop with quartic interactions :} Consider the ($++$) contribution to the trispectrum in the diagram of Fig.~\ref{Fig:Diagram_2}.
\\
Once again let us denote the magnitude of the momentum exchanged between the two sites as $s=|\vec{p}_1+\vec{p}_2|=|\vec{k}_1+\vec{k}_2|=|\vec{k}_3+\vec{k}_4|$, and the energies flowing through the vertices as,
\begin{align}
    E_L=\omega_1+\omega_2+p_1+p_2 , \hspace{15pt} E_R=\omega_3+\omega_4+p_1+p_2  \,.
\end{align}
\\
Evaluation of time integrals gives partial and total energy poles as follows,
\begin{align}
     & \langle \phi_{\vec{k}_1} \phi_{\vec{k}_2} \phi_{\vec{k}_3} \phi_{\vec{k}_4} \rangle^{1-\text{loop}} \sim \frac{g_4^2}{\omega_T}  \int d^d \vec{p}_1 d^d \vec{p}_2 ~\frac{1}{p_1 p_2} \frac{1}{E_L E_R} \left( \omega_T+2 p_+ \right) \,.
 \end{align}
 \\
 Upon performing the loop integrals both partial energy poles once again evaluate to branch cuts as follows,
\begin{align}
    &  \langle \phi_{\vec{k}_1} \phi_{\vec{k}_2} \phi_{\vec{k}_3} \phi_{\vec{k}_4} \rangle^{1-\text{loop}} \sim - \frac{g_4^2}{\omega_T} \left( \frac{1}{\delta} + \frac{1}{2} \log \left( \frac{\left( \omega_{12}+s\right) \left(\omega_{34}+s \right)}{\mu^2} \right) \right) +\text{NLf} \,,
\end{align}
where $\omega_{12}=\omega_1+\omega_2$ and $\omega_{34}=\omega_3+\omega_4$.
\\
Note that in both diagrams of Fig.~\ref{Fig:Diagram_1} and \ref{Fig:Diagram_2} the only subgraphs are trivial (i.e. the full graph) and only involve the loop. Let us now consider diagrams which have non-trivial subgraphs. \\

\begin{figure}[h]
\centering
\includegraphics[scale=0.5]{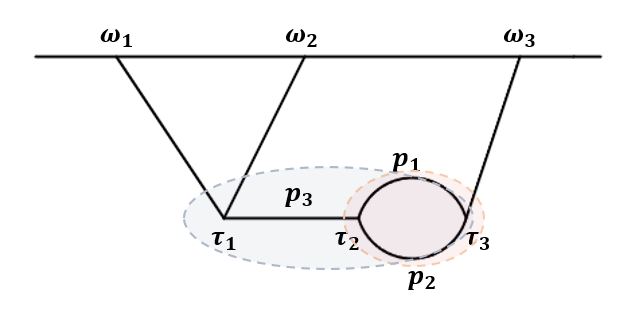}
\caption{Bispectrum at 1-loop correction with interaction $g_3\phi^3$. The blue-shaded subgraph is trivial. The red-shaded part denotes the subgraph of this diagram only involving the loop.}
\label{Fig:Diagram_3}
\end{figure}

\noindent \textbf{Bispectrum at 1-loop with three cubic interactions :} Consider the ($+++$) contribution to the bispectrum in the diagram of Fig.~\ref{Fig:Diagram_3}. Notice there are three subgraphs to this diagram, one of which is trivial. Two of the subgraphs involve the loop, including the trivial one, and are shaded in blue and red. The magnitude of the momentum flowing between loop sites is $s=|\vec{p}_1+\vec{p}_2|$. The energy flowing through the vertices $\tau_1, \tau_2$ and $\tau_3$ are $E_1, E_2$ and $E_3$ respectively. The partial energies flowing through subgraphs of this diagram are given by,
\begin{align}
    & E_1=\omega_1+\omega_2+p_3 , \hspace{15pt} E_2=p_3+p_1+p_2 , \hspace{15pt} E_3=p_1+p_2+\omega_3 , \nonumber \\ 
    & E_L=\omega_1+\omega_2+p_1+p_2 , \hspace{15pt} E_R=\omega_3+p_3 \,. \label{eqn:partial_energy_1}
\end{align}
\\
Evaluation of the time integrals leads to poles in total energy and partial energies as follows,

 \begin{align}
 & \langle \phi_{\vec{k}_1} \phi_{\vec{k}_2} \phi_{\vec{k}_3} \rangle^{1-\text{loop}} \sim  \frac{g_3^3}{p_3 E_1 E_R \omega_T}  \int d^d \vec{p}_1 d^d \vec{p}_2 ~\frac{1}{p_1 p_2}  \frac{1}{ E_2  E_3 E_L} \bigg[2 p_3^2 \left(\omega_T+2 p_+\right) \nonumber \\
     & +(\omega_{12}+p_+) (\omega_3 +2p_+) \omega_T  +p_3 \bigg(\omega_1 \left(2 \omega_2+3 \omega_3+5 p_+\right)+5 \omega_3 p_1+p_2 \left(5 \omega_3+8 p_1\right) \nonumber \\ 
     & +\omega_2 \left(3 \omega_3+5 p_+\right)+\omega_1^2+\omega_2^2+2 \omega_3^2+4 p_1^2+4 p_2^2 \bigg) \bigg] \,.
 \end{align}
\\
The poles in total energy $\omega_T$ and partial energies $E_1,E_R$ are independent of loop momenta and hence are pulled out of the integrals. After performing the loop integrals the partial energy poles in $E_2,E_3$ and $E_L$ once again evaluate to logarithms as follows,
\begin{align}
    &  \langle \phi_{\vec{k}_1} \phi_{\vec{k}_2} \phi_{\vec{k}_3} \rangle^{1-\text{loop}} \sim  \frac{g_3^3}{p_3 E_1 E_R \omega_T}(\omega_T+2p_3) \left[ \frac{1}{\delta} + \frac{1}{2} \log \frac{\left( \omega_3 +s \right) \left( p_3+s \right) }{\mu^2} \right] \nonumber \\
    & + \frac{g_3^3}{\omega_T E_1 E_R} \frac{\omega_3 +p_3}{\omega_{12}-p_3} \log \left(\frac{\omega_{12} +s}{p_3+s}\right)+\text{NLf}  \,, \label{Eqn:3ptcubic}
\end{align}
\\
where in the second line there is no folded pole at $\omega_{12}=p_3$. 

Of course the \textit{on-shell correlator} can be found from Eqn.~\eqref{Eqn:3ptcubic} by putting the external legs onshell, i.e. $\omega_i=k_i$ and using $s=p_3 =|\vec{k}_3|=\omega_3$, which simplifies the previous expression to,
\begin{align}
    &  \langle \phi_{\vec{k}_1} \phi_{\vec{k}_2} \phi_{\vec{k}_3} \rangle^{1-\text{loop}} \sim  \frac{g_3^3}{k_3 E_1 E_R k_T}\bigg( (k_T+2k_3) \left[ \frac{1}{\delta} + \frac{1}{2} \log \frac{k_3 E_R }{\mu^2} \right] \nonumber \\
    & + \frac{2 k_3^2}{k_{12}-k_3} \log \left(\frac{k_T}{E_R}\right) \bigg) +\text{NLf} \,.
\end{align}\\

\begin{figure} [h]
\centering
\includegraphics[scale=0.55]{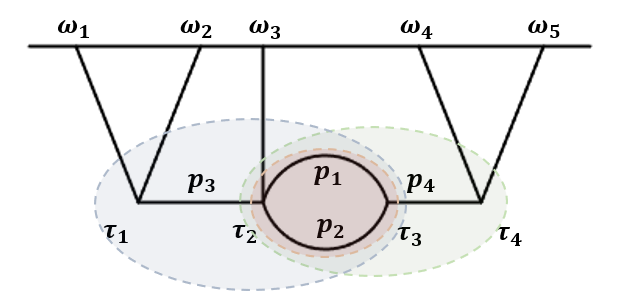}
\caption{5-point function at 1-loop with interactions $g_3 \phi^3$ and $g_4 \phi^4$. The red, green and blue shaded parts denote non-trivial subgraphs of this diagram involving the loop. The red shaded subgraph involves only the loop.}
\label{Fig:Diagram_4}
\end{figure}

\noindent \textbf{5-point function at 1-loop with cubic and quartic interactions :} Finally let us consider the ($++++$) contribution to the 5-point function in the diagram of Fig.~\ref{Fig:Diagram_4}. Notice there are six subgraphs for this diagram, one of which is trivial. Four of these involve the loop, including the trivial one, and the non-trivial ones are shaded in red, blue and green in Fig.~\ref{Fig:Diagram_4}. The red-shaded subgraph involves only the loop.
\\
Here the partial energies entering vertices and subgraphs of the diagram are,
\begin{align}
    & E_1=\omega_1+\omega_2+p_3, \hspace{15pt} E_2=\omega_3+p_3+p_1+p_2, \hspace{15pt} E_3=p_1+p_2+p_4, \hspace{15pt} E_4=p_4+\omega_4+\omega_5, \nonumber \\ 
    & E_{L_1}=\omega_1+\omega_2+\omega_3+p_4, \hspace{15pt} 
    E_{L}=\omega_1+\omega_2+\omega_3+p_1+p_2, \hspace{15pt} E_{M}=p_3+\omega_3+p_4, \nonumber \\
    & E_{R}=p_1+p_2+\omega_4+\omega_5, \hspace{15pt}
    E_{R_1}=p_3+\omega_3+\omega_4+\omega_5  \,.
    \label{Eqn:5-pt_1loop}
\end{align}
\\
The time integrals produce poles in total energy and partial energies. Once again the partial energies depending on loop momenta (i.e. $E_2, E_3, E_L$ and $E_R$) turn into logarithms once loop integrals are evaluated, giving us the following,
\begin{align}
   &\langle \phi_{\vec{k}_1}... \phi_{\vec{k}_5} \rangle^{1-\text{loop}} \sim g_4 g_3^3
   \bigg\{ \frac{2}{\delta}\frac{1}{\omega_T E_1p_4} \left( \frac{-1}{E_{L_1}} -\frac{2 E_1+E_M}{E_{L_1} E_M} -2\frac{E_1}{E_M E_{R_1}} - 2\frac{E_1+E_{R_1}}{E_{R_1} E_4} \right) \nonumber \\
   &+ 4 \frac{E_1+E_{R_1}}{\omega_T E_1 E_{R_1} E_4 } \frac{ \log \left[ \left(\omega_{45}+s \right) / \mu \right]}{\omega_{45}-p_4 } -2 \frac{E_1+E_M}{p_4 E_1 E_{L_1} E_M } \frac{\log \left[ \left( p_4+s \right) / \mu \right]}{\omega_{45}-p_4} \nonumber \\
   &  -2 \frac{E_M+E_4}{p_4 E_{R_1} E_4 E_M} \frac{\log \left[ \left( \omega_3+p_3+s \right)/ \mu \right]}{\omega_{12}-p_3} +4 p_3 \frac{E_{L_1}+E_4}{ \omega_T p_4 E_1 E_{L_1} E_4} \frac{\log \left[ \left( \omega_{123}+s \right) /\mu \right]}{\omega
   _{12}-p_3} \bigg\} +\text{NLf}\,,
\end{align}
where the correlator above is finite in the limit $\omega_{12}\rightarrow p_3$ and $\omega_{45}\rightarrow p_4$, and hence there are no folded poles.

\subsection{Correlators in de-Sitter}
\label{Sec:de-Sitter examples}
When the internal and external fields are either massless or conformally coupled scalars, we can write the correlator as a derivative of flat space expressions~\cite{Hillman:2021bnk,Lee:2023jby}, with one important caveat: since the de-Sitter mode function are $d$-dimension, they contribute additional logarithmic terms. 
We will expand on this point in Section~\ref{Sec:Loophole}.\\

\noindent \textbf{Bispectrum at 1-loop with cubic interactions:} As a simple example let us consider the 1-loop corrections to the 3-pt function in Fig.~\ref{Fig:Diagram_3} with the vertices ordered as $(+++)$, along with the contributions $(-++),(--+),(+-+)$. The left vertex is mediated by a $\lambda_1\dot{\phi}^3$ interaction while the middle and right vertex is mediated by the interaction $\lambda_2\dot{\phi}\sigma^2$. Hence the external lines are all massless scalars, the internal line with momentum $p_3$ is a massless scalar and the scalars in the loop are conformally coupled scalars. 

The correlators is given by:
\begin{align}
    \nonumber\langle \phi_{\vec{k}_1} \phi_{\vec{k}_2} \phi_{\vec{k}_3} \rangle^{1-\text{loop}}_{dS}&=\sum_{a_{L,M,R}=\pm}\frac{i^3a_La_Ra_M\lambda_1\lambda_2^2}{8k_1^3k_2^3k_3^3}\\&\times\int_{-\infty}^{0}\frac{d\tau_L}{\tau_L^{4+\delta}}\frac{d\tau_M}{\tau_M^{4+\delta}}\frac{d\tau_R}{\tau_R^{4+\delta}}\int\frac{d^{3+\delta}p}{(2\pi)^{3}}k_1^2k_2^2\tau_L^{4+\delta}k_3^2\tau_R^{2+\frac{\delta}{2}} e^{ia_L\omega_L\tau_L}e^{ia_L\omega_R\tau_R} \nonumber\\&\times\tau_L\tau_M\partial_{\tau_L}\partial_{\tau_M} G^{\phi}_{a_La_M}(p_3;\tau_L,\tau_M)G^{\sigma}_{a_Ma_R}(p_1;\tau_M,\tau_R)G^{\sigma}_{a_Ma_R}(p_2;\tau_M,\tau_R). \label{eqn:dSfigtimeintegral}
\end{align}
where $G$ is the propagator defined in Eqn.~\eqref{Eqn: Propagators}, $\omega_L=\omega_{12}$ and $\omega_R=\omega_3$. 

From the time integral representation \eqref{eqn:dSfigtimeintegral}, it is straightforward to see that the off-shell correlator is given by:
\begin{align}
    \langle \phi_{\vec{k}_1} \phi_{\vec{k}_2} \phi_{\vec{k}_3} \rangle^{1-\text{loop}}_{dS}(\omega_L,\omega_R) =-\bigg(1-\frac{3}{2}\delta\log(\frac{\omega_T}{H})\bigg)\partial_{\omega_L}^2\left(\langle \phi_{\vec{k}_1} \phi_{\vec{k}_2} \phi_{\vec{k}_3} \rangle^{1-\text{loop}}_{Flat}(\omega_L,\omega_R)\right) \,.
\end{align}

Here $\langle \phi_{\vec{k}_1} \phi_{\vec{k}_2} \phi_{\vec{k}_3} \rangle^{1-\text{loop}}_{Flat}(\omega_L,\omega_R)$ refers to the same diagram computed in flat space.

In this example, the \textit{unrenormalized} correlator gets an additional logarithmic correction. These logarithmic terms come expanding the mode function as well as the integration measure in the dim reg parameter $\delta$ (see \eqref{Eqn:Massless_d_dimension} and \eqref{measure_d}). There are 3 time integrals (each contributing $-\delta\log(-H\tau)$) and 9 mode function (3 from external mode functions and 6 from internal propagators, each contributing $\frac{\delta}{2}\log(-H\tau)$), and so upon time integration this gives a contribution of $-\frac{3\delta}{2}\log(\omega_T)$. This will be expanded further in section \ref{Sec:working_principle_deSitter}. 

A straightforward calculation leads us to the following expression:
\begin{align}
    & \langle \phi_{\vec{k}_1} \phi_{\vec{k}_2} \phi_{\vec{k}_3} \rangle^{1-\text{loop}}_{dS} \equiv \lambda_1\lambda_2^2 \bigg[ \frac{p_3^2 \left(\omega_R+3 \omega _L\right)+p_3 \left(3 \omega_R \omega _L+\omega_R^2+3 \omega _L^2\right)+p_3^3+\omega _T^3}{E_1^3 E_R \omega _T^3}\left(\frac{1}{\delta}-\frac{3}{2}\log(\frac{\omega_T}{H})\right)\nonumber \\&-\frac{ \omega_R p_3 \left(p_3^4 \left(k_3^2+3 \omega _L^2\right)-p_3^2 \left(9 \omega _L^4-k_3^4\right)-9 k_3^2 \omega _L^4+3 k_3^4 \omega _L^2+10 \omega _L^6\right)}{E_1^3 \omega _T^3 \left(\omega _L-\omega_R\right){}^3 \left(p_3-\omega _L\right){}^3}\log \left(\omega _L+s\right) \nonumber \\&-\frac{p_3 \omega _L \left(3 \omega_R^2+\omega _L^2\right) }{\omega_T^3 E_R \left(p_3-\omega_R\right) \left(\omega_R-\omega _L\right){}^3}\log \left(\omega_R+s\right)+\frac{\omega_R \left(\omega _L^3+3 p_3^2 \omega_L\right)}{E_R E_1^3 \left(p_3-\omega_R\right) \left(p_3-\omega _L\right){}^3}\log \left(p_3+s\right) \bigg]  \,.\label{eqn:dS3ptunren}
\end{align}

Just like the flat space example, the folded singularities are spurious.

We would like to renormalize this divergent loop contribution. The contribution from the counterterm comes in the form of a 3-point exchange diagram (where the loop in Fig.~\ref{Fig:Diagram_3} collapses into a vertex). In this diagram the left vertex is mediated by a $\lambda_1\dot{\phi}^3$ interaction while the right is mediated by a $-\frac{\lambda_2^2}{\delta}\dot{\phi}^2$ interaction\footnote{Since we are considering an \textit{off-shell} diagram, it makes sense to talk about an exchange diagram where one of the vertex has a quadratic vertex. The \textit{on-shell} counterpart of the story is that figure \ref{Fig:Diagram_3} is a contact diagram where an external leg receives loop correction, and this exchange diagram with a quadratic vertex is now a contact diagram removing the divergence from this corrected propagator.}.

The exchange diagram reads:
\begin{align}
    \langle \phi_{\vec{k}_1} \phi_{\vec{k}_2} \phi_{\vec{k}_3} \rangle^{CT}_{dS}=\sum_{a_{L,R}=\pm}\frac{-a_La_Ri^2\lambda_1\lambda_2^2}{8k_1^3k_2^3k_3^3\delta}i\int_{-\infty}^{0}&\frac{d\tau_L}{\tau_L^{4+\delta}}\frac{d\tau_R}{\tau_R^{4+\delta}}k_1^2k_2^2\tau_L^{4+\delta}k_3^2\tau_R^{2+\frac{\delta}{2}}\nonumber\\&\times\tau_L\tau_R\partial_{\tau_L}\partial_{\tau_R}G_{a_La_R}(p_3;\tau_L,\tau_R) \,.
\end{align}
Evaluating this diagram after expanding the mode function and time integration measure in $\delta$ gives:
\begin{equation}
     \langle \phi_{\vec{k}_1} \phi_{\vec{k}_2} \phi_{\vec{k}_3} \rangle^{CT}_{dS}=-\lambda_1\lambda_2^2\left[\frac{p_3^3+\omega_T^3+p_3^2(3\omega_L+\omega_R)+p_3(3\omega_L^2+3\omega_L\omega_R+\omega_R^2)}{64k_1k_2k_3\omega_T^3p_3E_L^3E_R}\right]\left(\frac{1}{\delta}-\frac{1}{2}\log(\frac{\omega_T}{H})\right) \,.
\end{equation}

The renormalized correlator reads:
\begin{align}
    & \langle \phi_{\vec{k}_1} \phi_{\vec{k}_2} \phi_{\vec{k}_3} \rangle^{1-\text{loop}}_{dS}\sim \lambda_1\lambda_2^2 \bigg[-\frac{p_3 \omega _L \left(3 \omega_R^2+\omega _L^2\right) }{\omega_T^3 E_R \left(p_3-\omega_R\right) \left(\omega_R-\omega _L\right){}^3}\log \left(\frac{\omega_R+s}{\omega_T}\right)\nonumber \\& -\frac{ \omega_R p_3 \left(p_3^4 \left(\omega_R^2+3 \omega _L^2\right)-p_3^2 \left(9 \omega _L^4-\omega_R^4\right)-9 \omega_R^2 \omega _L^4+3 \omega_R^4 \omega _L^2+10 \omega _L^6\right)}{E_1^3 \omega _T^3 \left(\omega _L-\omega_R\right){}^3 \left(p_3-\omega _L\right){}^3}\log \left(\frac{\omega _L+s}{\omega_T}\right) \nonumber \\&+\frac{\omega_R \left(\omega _L^3+3 p_3^2 \omega_L\right)}{E_R E_1^3 \left(p_3-\omega_R\right) \left(p_3-\omega _L\right){}^3}\log \left(\frac{p_3+s}{\omega_T}\right) \bigg]  \,. \label{Eqn:ren_dS_example}
\end{align}
Notice that the arguments of the logarithmic terms are scale invariant, i.e. if we scale the energies by $\lambda$ as $\omega\rightarrow \lambda \omega, p_3\rightarrow \lambda p_3$, the argument remains unchanged. \\

\begin{figure} [h]
\centering
\includegraphics[scale=0.5]{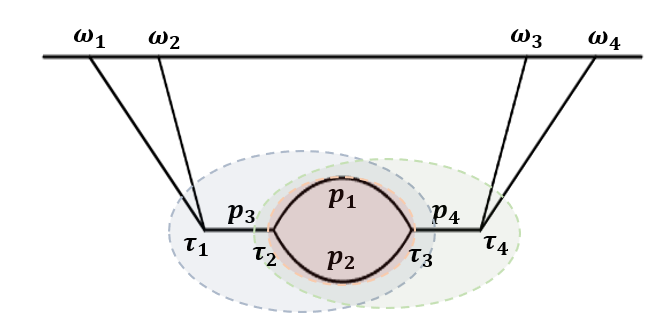}
\caption{1-loop correction to the 4-point function with cubic interactions $\lambda_1 \dot{\phi^3}$ and $\lambda_2 \dot{\phi} \sigma^2$. The shaded subgraphs involve the loop.}
\label{Fig:4_pt_conf}
\end{figure}

\noindent \textbf{4-point function at 1-loop with cubic interactions:} Once again considering interactions $\lambda_1 \dot{\phi^3}$ and $\lambda_2 \dot{\phi} \sigma^2$ with the conformally coupled scalars running in the loop and massless scalar running in all other legs, we consider the $(++++)$ contribution in Fig.~\ref{Fig:4_pt_conf} to the 4-point function at 1-loop.

Similar to the previous example, it is simple to see that the off-shell correlator in dS is related to the same diagram computed in flat space as the following,
\begin{align}
    \langle \phi_{\vec{k}_1} \phi_{\vec{k}_2} \phi_{\vec{k}_3} \phi_{\vec{k}_4} \rangle^{1-\text{loop}}_{dS}(\omega_L,\omega_R) =\bigg(1-2 \delta\log\left(\frac{\omega_T}{H}\right)\bigg) \partial_{\omega_L}^2 \partial_{\omega_R}^2\left(\langle \phi_{\vec{k}_1} \phi_{\vec{k}_2} \phi_{\vec{k}_3}\phi_{\vec{k}_4} \rangle^{1-\text{loop}}_{Flat}(\omega_L,\omega_R)\right) \,,
\end{align}
where $\omega_L=\omega_1+\omega_2$ and $\omega_R=\omega_3+\omega_4$ are the external energies flowing into the left and right vertices $\tau_1$ and $\tau_4$ respectively. The \textit{unrenormalized} correlator gets an additional logarithmic correction of $-2\delta \log\big(\frac{\omega_T}{H}\big)$ from 4 vertices (each contributing $-\delta \log \big(-H\tau\big)$) and 12 modes (each contributes $\frac{\delta}{2}\log \big(-H\tau \big)$). 

Again, similar to the previous example, renormalizing this loop diagram would give contributions from 4-point exchange diagram, with the loop collapsing to a vertex mediated by a dimension-4 quadratic counter term, e.g. $\frac{\lambda_2^2}{\delta} \dot{\phi}^2$. 

After a straightforward computation, the final renormalized 4-point correlator reads,

\begin{align}
    &\langle \phi_{\vec{k}_1} \phi_{\vec{k}_2} \phi_{\vec{k}_3} \phi_{\vec{k}_4} \rangle^{1-\text{loop}}_{dS} \sim \lambda_1^2 \lambda_2^2 \bigg[\frac{-2 \log \left(\frac{\omega_L + s}{\omega_T} \right)}{\omega _T^5 \left(p_3-\omega _L\right){}^3 \left(\omega _L+p_3\right){}^3 \left(\omega _L+p_4\right){}^3 \left(p_4+\omega _R\right){}^3} Q_1(\omega_L,\omega_R,p_3,p_4) \nonumber \\
    & + \frac{ \log \left(\frac{p_4+s}{\omega_T}\right)}{\left(p_3+p_4\right) \left(\omega _L+p_3\right){}^3 \left(\omega _L+p_4\right){}^3 \bigg(p_4-\omega _R\bigg){}^3}\bigg(2 p_3^2 \left(3 \omega _L+p_4\right)+2 p_3 \left(3 p_4 \omega _L+3 \omega _L^2+p_4^2\right)\nonumber \\
    & +\left(\omega _L+p_4\right){}^3+2 p_3^3\bigg) -\frac{2 \log\left( \frac{s + \omega_R}{\omega_T}\right)}{\omega _T^5 \left(\omega _L+p_3\right){}^3 \left(p_3+\omega _R\right){}^3 \left(p_4^2-\omega _R^2\right){}^3} Q_2(\omega_L,\omega_R,p_3,p_4)
    \nonumber \\
    & +\frac{ \log \left(\frac{p_3+s}{\omega_T}\right)}{\left(p_3-p_4\right) \left(p_3-\omega _L\right){}^3}\left(\frac{1}{\left(p_4+\omega _R\right){}^3}-\frac{2 p_4}{\left(p_3+p_4\right) \left(p_3+\omega _R\right){}^3}\right)\bigg] 
\end{align}
where $Q_1$ and $Q_2$ are polynomials of external and exchange energies, given by,
\begin{align}
        &Q_1=p_3 \bigg((2 p_4 \omega _T^3 \left(p_3^2 \left(p_4^2-3 \omega _L^2\right)+\omega _L^2 \left(8 p_4 \omega _L+6 \omega _L^2+3 p_4^2\right)+p_3^4\right) \left(p_4+\omega _R\right) +2 p_4 \omega _T^2 \big(p_3^2 \left(p_4^2-3 \omega _L^2\right)\nonumber \\
    & +\omega _L^2 \left(8 p_4 \omega _L+6 \omega _L^2+3 p_4^2\right)+p_3^4\big) \left(p_4+\omega _R\right){}^2+2 p_4 \omega _T \left(\omega _L+p_4\right) \left(p_3^2-\omega _L^2\right)\left(p_3^2-\omega _L \left(3 \omega _L+2 p_4\right)\right) \nonumber \\
    & \left(2 p_4 \left(\omega _L+4 \omega _R\right)+4 \omega _L \omega _R+\omega _L^2+3 p_4^2+6 \omega _R^2\right)+2 p_4 \left(\omega _L+p_4\right){}^2 \left(p_3^2-\omega _L^2\right){}^2 \big(3 p_4 \left(\omega _L+5 \omega _R\right)\nonumber \\
    &+5 \omega _L \omega _R+\omega _L^2+6 p_4^2+10 \omega _R^2\big)+\omega _T^4 \left(p_3^2 \left(p_4^2-3 \omega _L^2\right)+\omega _L^2 \left(8 p_4 \omega _L+6 \omega _L^2+3 p_4^2\right)+p_3^4\right) \left(2 p_4+\omega _T\right)\bigg), \nonumber 
\end{align}
\begin{align}
    & Q_2= p_4 \bigg(6 p_3^4 \omega _T \left(p_4^2 \left(-4 \omega _L \omega _R+\omega _L^2-17 \omega _R^2\right)+3 \omega _R^2 \left(4 \omega _L \omega _R+\omega _L^2+5 \omega _R^2\right)+6 p_4^4\right)\nonumber \\
        & +p_3^2 \omega _T \big(18 p_4^4 \left(4 \omega _L \omega _R+\omega _L^2+\omega _R^2\right)+p_4^2 \left(4 \omega _L^3 \omega _R-84 \omega _L^2 \omega _R^2-212 \omega _L \omega _R^3+\omega _L^4-53 \omega _R^4\right) \nonumber \\
        & +3 \omega _R^2 \left(20 \omega _L^3 \omega _R+70 \omega _L^2 \omega _R^2+84 \omega _L \omega _R^3+\omega _L^4+21 \omega _R^4\right)\big)+2 p_3^5 \big( p_4^2 \left(-4 \omega _L \omega _R+\omega _L^2-17 \omega _R^2\right)\nonumber \\
        & +3 \omega _R^2 \left(4 \omega _L \omega _R+\omega _L^2+5 \omega _R^2\right)+6 p_4^4\big)+2 p_3^3 \big(p_4^4 \left(50 \omega _L \omega _R+19 \omega _L^2+19 \omega _R^2\right)+3 p_4^2 \left(\omega _L+2 \omega _R\right) \left(\omega _L+3 \omega _R\right) \nonumber \\
        & \left(-6 \omega _L \omega _R+\omega _L^2-3 \omega _R^2\right)+3 \omega _R^2 \left(21 \omega _L^3 \omega _R+54 \omega _L^2 \omega _R^2+57 \omega _L \omega _R^3+3 \omega _L^4+17 \omega _R^4\right)\big)\nonumber \\
        & +2 p_3 \big(3 p_4^4 \left(5 \omega _L^3 \omega _R+10 \omega _L^2 \omega _R^2+5 \omega _L \omega _R^3+\omega _L^4+\omega _R^4\right)-3 p_4^2 \omega _R^2 \left(17 \omega _L^3 \omega _R+30 \omega _L^2 \omega _R^2+15 \omega _L \omega _R^3+3 \omega _L^4+3 \omega _R^4\right)\nonumber \\
        &+2 \omega _R^3 \left(19 \omega _L^4 \omega _R+56 \omega _L^3 \omega _R^2+70 \omega _L^2 \omega _R^3+35 \omega _L \omega _R^4+2 \omega _L^5+7 \omega _R^5\right)\big)+\omega _T^5 \left(-3 p_4^2 \omega _R^2+p_4^4+6 \omega _R^4\right)\bigg)
\end{align}

Once again notice that the logarithmic terms are scale invariant, and like the flat space examples all apparent folded poles are spurious.

\section{Singularity Structure of Loop Corrections}

The singularity structure for tree level correlators has been well studied in the literature \cite{Baumann:2021fxj,Baumann:2020dch,Arkani-Hamed:2017fdk,Arkani-Hamed:2018bjr,Benincasa:2018ssx}. It is well known that correlation functions of massless fields at tree level have singularities of total and partial (energy flowing into any subgraph) energy pole. The residue at a partial energy pole is given by the product of a lower-point correlator and a lower-point scattering amplitude \cite{Baumann:2020dch,Arkani-Hamed:2017fdk,Arkani-Hamed:2018bjr,Benincasa:2018ssx,Goodhew:2020hob}, whereas the residue at the total energy singularity is related to the scattering amplitude for the same process \cite{Maldacena:2011nz,Raju:2012zr}.\\
The singularity structure of correlators at loops is comparatively much more complicated. Some recent progress has been made in studying the types of singularities arising in wavefunction coefficients at 1-loop (see \cite{Benincasa:2024ptf,Chowdhury:2023khl,Beneke:2023wmt,Chowdhury:2023arc}). The singularity structure of correlators can however be very different due to different boundary conditions in the past, and as a result different bulk-bulk propagators. Thus correlators at loop level might have additional singularities or some singularities might cancel. \\ 
Now, in dimensional regularisation it is convenient to perform the time integrals first, which produces total and partial energy poles. Some of these partial energy poles will involve loop momenta. It is interesting to investigate what kind of singularities such poles produce once momentum integrals are performed. Importantly, there is a way to know the singularity structure of these diagrams without any explicit calculation. We will discuss the diagrammatic rules to extract singularities (the poles and branch cuts) in a correlator at 1-loop in Sec.~\ref{Sec:Singularity_structure}, and in Sec.~\ref{Sec:Working Principle} we will demonstrate how these rules follow from the general structure of the integrals in these correlators.

\subsection{Rules to extract Singularity Structure at 1-loop} \label{Sec:Singularity_structure}
There are a few lessons to be learnt from the results of Sec.~\ref{Sec:Explicit_Comp}. We notice that the singularity structure of the correlators at 1-loop feature poles and logarithms. These poles in total and partial energies (ofcourse, independent of loop momenta) entering vertices and subgraphs of the diagram are produced from time integrals, and also arise in tree level correlations of massless scalars. The branch cuts however are produced from momentum integrals (in dS one also gets logarithms of total energy from time integrals, see Sec.~\ref{Sec:Renormalization}). Regarding the branch cuts produced, we see a trend emerging, as follows, \\
\begin{itemize}
    \item First, we notice the obvious fact that all diagrams have at least one subgraph, trivial or otherwise, which involves only the loop. We call these subgraphs as \textbf{Loop-subgraphs}. While the \textit{loop subgraph} is trivial for Figs.~\ref{Fig:Diagram_1} and \ref{Fig:Diagram_2}, it has been shaded in red in Figs.~\ref{Fig:Diagram_3}, \ref{Fig:Diagram_4} and \ref{Fig:4_pt_conf}. These subgraphs have two vertices of \textit{external} energy injection, through which the energy flows \textit{into} the subgraph. These vertices are of course the loop sites. For example the energy flowing into the \textit{loop-subgraph} through the left loop site in the diagrams of Figs.~\ref{Fig:Diagram_1}, \ref{Fig:Diagram_2} is $\mathcal{S}_L=\omega_1+\omega_2$, in Figs. \ref{Fig:Diagram_3}, \ref{Fig:4_pt_conf} is $\mathcal{S}_L=p_3$ and in Fig. \ref{Fig:Diagram_4} is $\mathcal{S}_L=p_3+\omega_3$. The energy flowing in through the right loop site in Figs.~\ref{Fig:Diagram_1},~\ref{Fig:Diagram_3} is $\mathcal{S}_R=\omega_3$, in Fig.~\ref{Fig:Diagram_2} is $\mathcal{S}_L=\omega_3+\omega_4$, and in Figs. \ref{Fig:Diagram_4},~\ref{Fig:4_pt_conf} is $\mathcal{S}_L=p_4$. 
    
    Importantly, note that all these diagrams have a branch cut at $\mathcal{S}_L +s$ and $\mathcal{S}_R +s$.

\item Apart from \textit{loop-subgraphs}, we considered diagrams in Figs.~\ref{Fig:Diagram_3}, \ref{Fig:Diagram_4} and \ref{Fig:4_pt_conf} which have additional subgraphs involving the full loop. For Fig. \ref{Fig:Diagram_3} this subgraph is a trivial one and shaded in blue. For Figs.~\ref{Fig:Diagram_4} and \ref{Fig:4_pt_conf} there are 3 such subgraphs and the non-trivial ones are shaded in green and blue. Notice that atleast one of the \textit{endpoints} of these subgraphs is a loop site, which means that atleast one of the loop sites is not connected to any other vertex of the subgraph, except the other loop site (via, ofcourse the loop arms). Any subgraph which does not involve the full loop, or which does not have a loop site as an endpoint will \textit{not} be considered for determining the logarithmic contributions to the singularity structure of loop diagrams. Note that the trivial subgraph in Figs.~\ref{Fig:Diagram_4} and \ref{Fig:4_pt_conf} are such examples, and as mentioned we will not be considering this subgraph. 

Now, the loop divides the entire diagram into two parts, one to the right of the loop and the other to the left. We call the subgraphs to the left and to the right of the loop, with one loop site as an \textit{endpoint}, as \textbf{left-subgraphs} and \textbf{right-subgraphs} respectively. Fig. \ref{Fig:Diagram_3} has one \textit{left-subgraph} (shaded in blue). Let us call the external energy injection into this subgraph through all the vertices except the right loop site as $\mathcal{S}_{L_1}=\omega_1+\omega_2$. Figs. \ref{Fig:Diagram_4} and \ref{Fig:4_pt_conf} have one \textit{left-subgraph} (shaded in blue) and one \textit{right-subgraph} (shaded in green). The external energy injection into the \textit{left-} (\textit{right-}) \textit{subgraph} through all the vertices except the right (left) loop site is given by $\mathcal{S}_{L_1}=\omega_1+\omega_2+\omega_3$ and $\mathcal{S}_{L_1}=\omega_1+\omega_2$ $\big( \mathcal{S}_{R_1}=\omega_4+\omega_5\big)$ for Figs. \ref{Fig:Diagram_4} and \ref{Fig:4_pt_conf} respectively.

Once again, note that Fig.~\ref{Fig:Diagram_3} has a branch cut at $\mathcal{S}_{L_1}+s$ and Figs.~\ref{Fig:Diagram_4} and \ref{Fig:4_pt_conf} have branch cuts at $\mathcal{S}_{L_1}+s$ and $\mathcal{S}_{R_1}+s$. \\
\end{itemize}

\noindent This leads us to the one of the main results of this work : \textbf{Rules to extract the singularity structure of any 1-loop diagram at 2-sites}, given below,

\begin{itemize}
\item The correlator will feature poles in total and partial energies, entering vertices and subgraphs of the diagram, which are independent of loop momenta $p_1,p_2$. These poles are produced from time integrals. 
    \item Consider the energy flowing into the \textit{loop-subgraphs} through the left and right loop sites. We call this energy as $\mathcal{S}_L$ and $\mathcal{S}_R$ respectively. The loop correction coming from the diagram will feature a logarithm of $\mathcal{S}_L+s$ and $\mathcal{S}_R+s$.    
    \item For every \textit{left-} and \textit{right- subgraph} consider the energy flowing into the subgraph through all vertices except the right and the left loop site respectively. Let us call these energies as $\mathcal{S}_{L_i}$ and $\mathcal{S}_{R_j}$ respectively. The loop correction coming from the diagram will feature a logarithm of $\mathcal{S}_{L_i} +s $ and $\mathcal{S}_{R_j} +s$.\\
\end{itemize} 
 The number of logarithms hence produced matches with the number of partial energies depending on loop momenta produced after time integrals, which should turn into branch cuts once loop integrals are performed. As we shall see in the following, these rules will apply for derivative as well as polynomial interactions in flat space. In de Sitter one requires derivative interactions if only massless scalars are involved, however, if conformally coupled scalars run in the loop then the rules are valid for both derivative as well as polynomial interactions. \\
Note that in all the diagrams considered above, the vertices lie on the same side of $\tau=0$. However, a correlator will have contributions coming from diagrams where the vertices will come from the time ordered as well as the anti-time ordered parts of the in-in expression. We discuss in Sec.~\ref{Sec:Unord} that the rules mentioned above still apply on such diagrams,  with an additional constraint that the vertices of the \textit{left-} and \textit{right-subgraphs} should lie on the same side of $\tau=0$. 

\subsection{Working Principle}
\label{Sec:Working Principle}
In the following sections, we will study the general structure of correlators and the singularities arising from these integrals, and we will see how the diagrammatic rules follow. 

\subsubsection{$n$-point functions in Flat Space} \label{Sec:working_principle_flatspace}

\begin{figure}[h]
\centering
\includegraphics[scale=0.55]{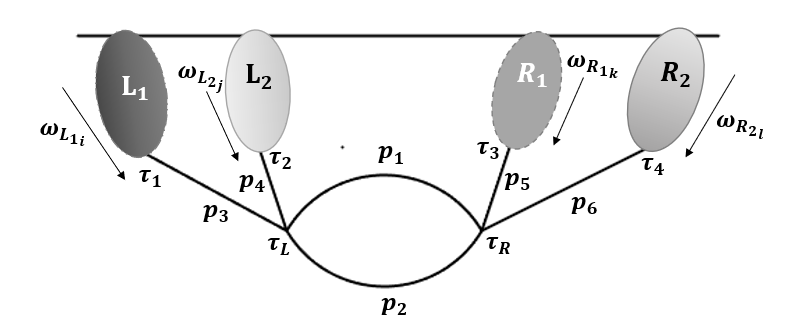}
\caption{A general 1-loop correction diagram. The blobs denote tree-level subdiagrams with $p_3,p_4,p_5$ and $p_6$ as momentum exchanges with the loop sites. The different color gradients of the blobs represent the fact that these subdiagrams may in general be different from each other.}
\label{Figure:general}
\end{figure}

Consider a general 1-loop diagram with quartic and cubic polynomial interactions in Fig.~\ref{Figure:general}. Loop momenta $p_1$ and $p_2$ flow between loop sites $\tau_R$ and $\tau_L$. Each loop site contracts with multiple legs ($p_3,...,p_6$), which may lead to further tree level sub-diagrams denoted by the blobs. The blobs to the left of the loop are labelled as $L_1$ and $L_2$ and the blobs to the right of the loop are labelled as $R_1$ and $R_2$. We refer to the energies flowing through the left blobs into the vertices $\tau_1$ and $\tau_2$ as $\omega_{{L_1}_i}$ and $\omega_{{L_2}_j}$ respectively, where $i,j$ run over partial energies flowing through all possible subgraphs of the blob, including the total energy flowing through the blob. Similarly the energies flowing through the right blobs into the vertices $\tau_3$ and $\tau_4$ are $\omega_{{R_1}_k}$ and $\omega_{{R_2}_l}$ respectively, with $k,l$ running over all possible partial energies.
\\
Evaluation of time integrals in this diagram will lead to simple poles in partial energies, and the ones involving $p_1$ and $p_2$ are given by, 
\begin{align}
    &  E_{{L_1}_i}=\omega_{{L_1}_i}+p_4+p_+, \hspace{15pt} E_{{L_2}_j}=\omega_{{L_2}_j}+p_3+p_+, \hspace{15pt} E_{{L}_{ij}}=\omega_{{L_1}_i}+\omega_{{L_2}_j}+p_+, \nonumber \\
    & E_{{R_1}_k}=\omega_{{R_1}_k}+p_6+p_+, \hspace{15pt} 
    E_{{R_2}_l}=\omega_{{R_2}_l}+p_5+p_+, \hspace{15pt} E_{{R}_{kl}}=\omega_{{R_1}_k}+\omega_{{R_2}_l}+p_+, \nonumber \\
    & E_1=p_3+p_4+p_+, \hspace{15pt} E_2=p_++p_5+p_6  \,, \label{Eqn:partial_general}
\end{align}
where $p_+=p_1+p_2$. After performing the time integrals, the loop correction has the following explicit dependence on loop momenta $p_1$ and $p_2$,
\begin{align}
    &\int d^d \vec{p}_1 d^d \vec{p}_2 ~\frac{Q(p_i,\omega_j)}{p_1 p_2 E_1 E_2 \prod_{ijkl} \left( E_{{L_1}_i} E_{{L_2}_j} E_{L_{ij}} E_{{R_1}_k} E_{{R_2}_l} E_{R_{kl}} \right) } \delta^d (\vec{p}_1 +\vec{p}_2 -\vec{s}) \nonumber \\
    & = \frac{S_{d-2}}{2} \int_{s}^{\infty} dp_+ \int_{-s}^{s} dp_- \frac{p_1 ^{d-2} p_2}{s} \sin^{d-3}{\theta_1}~\frac{Q(p_i,\omega_j)}{p_1 p_2 E_1 E_2 \prod_{ijkl} \left( E_{{L_1}_i} E_{{L_2}_j} E_{L_{ij}} E_{{R_1}_k} E_{{R_2}_l} E_{R_{kl}} \right) } \,, \label{Eqn:loop_gen}
\end{align}
where $\sin{\theta_1}=\sqrt{1-\left( \frac{s^2 +p_1^2-p_2^2}{2s p_1}\right)^2}$ and $p_-=p_1-p_2$. $s$ is the magnitude of the momentum exchanged between the loop sites, and $S_{d-2}$ is surface area of a $(d-2)$-unit sphere. $Q$ is a function of exchange momenta $p_i$ (excluding the loop arms) and $\omega_j$ (energy of the external legs). The equality in Eqn.~\ref{Eqn:loop_gen} follows from an identity (discussed in the Appendix of \cite{Bhowmick:2024kld}). Note that $Q$ may have simple poles in partial energies of subgraphs that do not depend on loop momenta (e.g. $\omega_{L_{1_i}}+p_3,~\omega_{R_{1_k}}+p_5$ and so on).

\begin{figure}[h]
\centering
\includegraphics[scale=0.45]{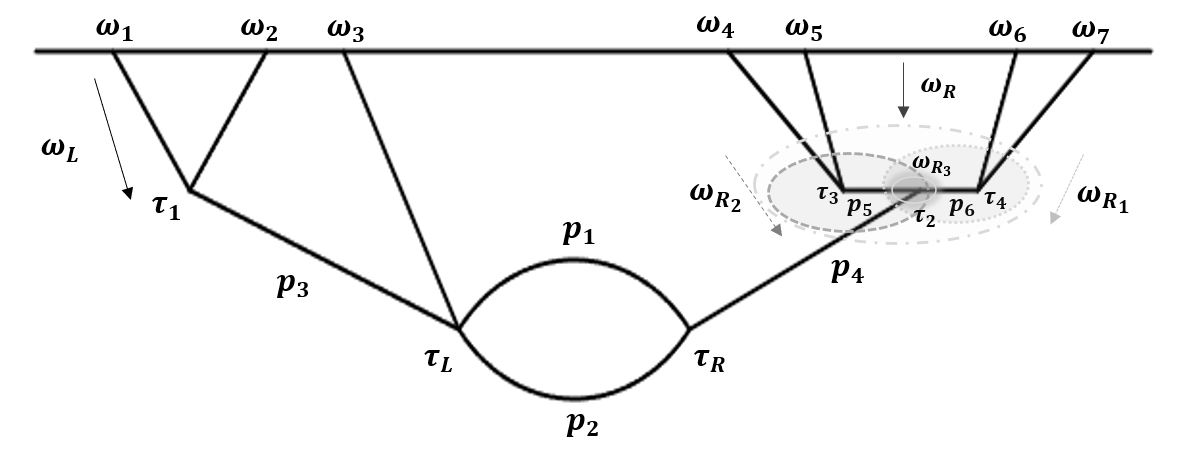}
\caption{A 1-loop correction diagram with cubic and quartic vertices. The grey blobs denotes various subgraphs.}
\label{Figure:general_example}
\end{figure}

Now, for the computation to be tractable, the integrand is scaled by $s$. With a slight abuse of notation by suppressing the product over $i,j,k,l$, we have the following dependence on $p_-$ and $p_+$,
\begin{align}
    & = s^{m+\delta} 2^{-\delta-1} S_{d-2} \int_{1}^{\infty} d\hat{p}_+ \int_{-1}^{1} d\hat{p}_-~\frac{Q(\hat{p}_i,\hat{\omega}_j)}{\hat{E}_{L_1} \hat{E}_{L_2} \hat{E}_L \hat{E}_{R_1} \hat{E}_{R_2} \hat{E}_R \hat{E}_1 \hat{E}_2} \left(\left(1-\hat{p}_-^2\right) \left(\hat{p}_+^2-1\right)\right){}^{\delta /2} \,, \label{Eqn:loopintegral_flat}
\end{align}
\\
where $m$ is an integer. The hat denotes scaled energies, i.e. $\hat{\omega}_j=\omega_j/s$ and so on. Since the only $p_-$ dependence in the integrand is the factor of $(1-\hat{p}_-^2)^{\delta/2}$, using $\int_{-1}^1 dp_- (1-\hat{p}_-^2)^{\delta/2}= \frac{\sqrt{\pi } \Gamma \left(\frac{\delta }{2}+1\right)}{\Gamma \left(\frac{\delta +3}{2}\right)}$, the above expression evaluates to the following,
\begin{align}
    & =s^{m+\delta} 2^{-\delta-1} S_{d-2} \frac{\sqrt{\pi } \Gamma \left(\frac{\delta }{2}+1\right)}{\Gamma \left(\frac{\delta +3}{2}\right)} \int_{1}^{\infty} d\hat{p}_+~\frac{Q(\hat{p}_i,\hat{\omega}_j)}{\hat{E}_{L_1} \hat{E}_{L_2} \hat{E}_L \hat{E}_{R_1} \hat{E}_{R_2} \hat{E}_R \hat{E}_1 \hat{E}_2}  \left(\hat{p}_+^2-1\right){}^{\delta /2} \,.
\end{align}
\\
Recall that the poles in the integrand are all simple, hence using partial fraction decomposition of the integrand in $p_+$ gives the following structure,
\begin{align}
    s^{\delta}\int_1^{\infty} d \hat{p}_+ \frac{(\hat{p}_+^2-1)^{\delta/2}}{\hat{p}_++x} = -\frac{1}{\delta }-\log (1+x)-\log (s)  \,, \label{Eqn:general_loop_1}
\end{align}
\\ where the $\hat{p}_++x$ in the denominator represents the various partial energies in Eqn.~\ref{Eqn:partial_general}. For example, upon partial fraction decomposition for the term with $\hat{E}_{{L_2}_j}$ in the denominator we have $x=\hat{\omega}_{{L_2}_j} +\hat{p}_3$ and the RHS of Eqn.~\ref{Eqn:general_loop_1} gives $-\frac{1}{\delta }-\log (s+\omega_{{L_2}_j} +p_3)$. Similarly for the term with $\hat{E}_1$ in the denominator we have $x=\hat{p}_3+\hat{p}_4$ and the RHS of Eqn.~\ref{Eqn:general_loop_1} gives $-\frac{1}{\delta }-\log (s+p_3+p_4)$, and so on.  
\\
From Eqn.~\ref{Eqn:general_loop_1} we see that every partial energy pole turns into a logarithm, and the argument of the logarithm is precisely the partial energy pole at $p_+ \rightarrow s$. The logarithms arising from the energies flowing through the loop sites $E_1$ and $E_2$ correspond to contributions from what we refer to as \textit{loop-subgraphs}. The logarithms arising from integrating the pole at partial energies flowing through blobs to the the left (right) of the loop, i.e. $E_{L_1}, E_{L_2}$ and $E_L$ ($E_{R_1},E_{R_2}$ and $E_R$) correspond to contributions from \textit{left-} (\textit{right-}) \textit{subgraphs}. To be precise, 
\begin{align}
    &E_1|_{p_+\rightarrow s}= \mathcal{S}_L~, \hspace{85pt}  E_2|_{p_+ \rightarrow s}= \mathcal{S}_R~,\nonumber \\  &E_{L_{(1i,~2j,~ij)}}|_{p_+\rightarrow s}= \mathcal{S}_{L_{(1i,~2j,~ij)}}~,\hspace{10pt} E_{R_{(1k,~2l,~kl)}}|_{p_+\rightarrow s}= \mathcal{S}_{R_{(1k,~2l,~kl)}} \,, \label{Eqn:bypart_diagrule}
\end{align}
where $\mathcal{S}_L,\mathcal{S}_R$ are the energies flowing into the \textit{loop-subgraph} through the left and right loop sites respectively. $\mathcal{S}_{L_{(1i,~2j,~ij)}}$ is the energy flowing into the \textit{left-subgraphs} (involving the blob $L_1, L_2$ and both respectively) through all vertices except the right loop site, and so on. This demonstrates that the rules mentioned in Sec.~\ref{Sec:working_principle_flatspace} are indeed the correct diagrammatic way of extracting the singularities. 
\\
To demonstrate how powerful this analysis is in extracting the singularity structure, we consider an explicit example of the previous arguments in the following. Consider the diagram in Fig.~\ref{Figure:general_example}. Comparing with Fig.~\ref{Figure:general} we see that there is only one left and right blob attached to the loop site. The energy flowing out of the left blob through the vertex $\tau_1$ is $\omega_L=\omega_1+\omega_2$. There are four partial energies (corresponding to the subgraphs in the grey-shaded regions in Fig.~\ref{Figure:general_example}) flowing out of the right blob through the vertex $\tau_2$, given by $\omega_{R_1}=\omega_6+\omega_7+p_5$, $\omega_{R_2}=\omega_4+\omega_5+p_6$, $\omega_{R_3}=p_5+p_6$ and $\omega_{R}=\omega_4+\omega_5+\omega_6+\omega_7$.
\\
Now to evaluate this diagram explicitly would require extremely lengthy time integrals (i.e. $6!=720$ time orderings) and complicated loop integrals. Evaluation of time integrals will lead to partial and total energy poles. The partial energy poles involving loop momenta are,
\begin{align}
     &  E_{L}=\omega_{{L}}+\omega_3+p_+, \hspace{15pt} E_1=p_3+\omega_3+p_+, \hspace{15pt} E_2=p_++p_4 \nonumber \\
    & E_{R_1}=\omega_{{R_1}}+p_+, \hspace{15pt} 
    E_{R_2}=\omega_{{R_2}}+p_+, \hspace{15pt} E_{R_3}=\omega_{{R_3}}+p_+, \hspace{15pt} E_R=\omega_R+p_+ \,.
    \label{Eqn:partial_general_example}
\end{align}
\begin{figure}[h]
\centering
\includegraphics[scale=0.58]{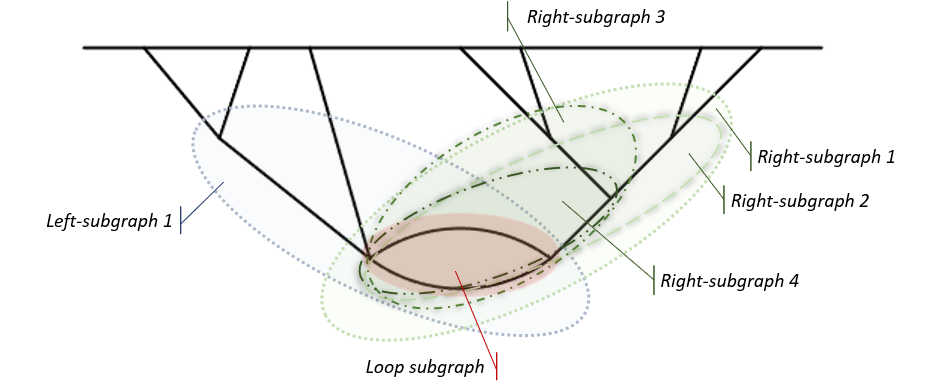}
\caption{The 1-loop correction diagram of Fig.~\ref{Figure:general_example} with the \textit{loop-subgraph} shaded in red, \textit{left-} and \textit{right- subgraphs} shaded in blue and green respectively. The momentum labels on the arms as well as vertices have been removed for the diagram to remain legible. The reader is requested to refer to Fig.~\ref{Figure:general_example} for these labels.}
\label{Figure:general_example_2}
\end{figure}
\\
Following the rules mentioned in Sec.~\ref{Sec:Singularity_structure} we will immediately know the singularity structure without any computations, as follows,
\begin{itemize}
    \item For the \textit{loop-subgraph} (shaded in red in Fig.~\ref{Figure:general_example_2}), we have $\log(\mathcal{S}_L+s)=\log (p_3+\omega_3+s)$ from the left loop site and $\log(\mathcal{S}_R+s)=\log (p_4+s)$ from the right loop site. 
    
    Note that this is precisely what we get from Eqn.~\ref{Eqn:general_loop_1} considering the $E_1$ and $E_2$ pole respectively in the partial fraction decomposition.
    
    \item From the \textit{left-subgraph} (shaded in blue in Fig.~\ref{Figure:general_example_2}), we have $\log(\mathcal{S}_{L_1}+s)=\log (\omega_1+\omega_2+\omega_3+s)$.

    Again note this is precisely what we get from Eqn.~\ref{Eqn:general_loop_1} considering the $E_{L}$ pole in the partial fraction decomposition.
    
    \item From the four \textit{right-subgraphs} (shaded in green in Fig.~\ref{Figure:general_example_2}), we have $\log(\mathcal{S}_{R_1}+s)=\log (\omega_6+\omega_7+p_5+s)$, $\log(\mathcal{S}_{R_2}+s)=\log (\omega_4+\omega_5+p_6+s)$, $\log(\mathcal{S}_{R_3}+s)=\log (p_5+p_6+s)$ and $\log(\mathcal{S}_{R_4}+s)=\log (\omega_4+\omega_5+\omega_6+\omega_7+s)$.

    Once again note this is precisely what we get from Eqn.~\ref{Eqn:general_loop_1} considering the $E_{R_1}$, $E_{R_2}$, $E_{R_3}$ and $E_R$ poles respectively in the partial fraction decomposition.
\end{itemize}

 To conclude, this shows how the diagrammatic rules follow from the general structure of a correlator in flat space with polynomial interactions. Notice that with interactions where derivative acts on the fields, such as $\dot{\phi}^3$ or $\big( \partial_i \phi \big)^2 \phi^2$ the general structure of the integral is still going to be as in Eqn~\eqref{Eqn:loop_gen}, except now the factor of $p_1p_2$ will feature in the numerator. In that case our arguments still go through and the integral looks like Eqn.~\eqref{Eqn:general_loop_2} with $\alpha=1$ (The value of $y$ obviously depends on the number of derivatives). 

 In the following we consider a general $n-$point function in de-Sitter and show how the diagrammatic rules apply to de-Sitter.

\subsubsection{$n$-point functions in de-Sitter} \label{Sec:working_principle_deSitter}

Unlike flat space, evaluation of time integrals in de-Sitter can produce higher order poles. Additionally the mode functions and measure are dependent on dimensions (Eqns.~\eqref{Eqn:massless_dS},~\eqref{measure_d}), but can be written as a sum of contribution from a three dimensional part, and a $\mathcal{O}(\delta)$ part. Thus the entire correlator may be expressed as a sum of contribution coming from the former and the latter, which we refer to as $D_1$ and $D_2$ respectively. In this section, we compute the general structure of $D_1$ to see what kind of singularities may arise. We will see why the rules to extract the singularities in flat space very much apply to de-Sitter as well. 
\\
Also, note that the integral of $D_2$ is the same as that of $D_1$ with an extra factor of $\log(-H \tau_i)$. Consequently, the result of performing the time integrals in $D_2$ is the same as that of $D_1$, with an extra factor of $\log (H/\omega_T)$, with some rational functions of momenta \cite{Bhowmick:2024kld}, i.e.,
\begin{align}
    D_2(\omega_i)=\delta~ D_1(\omega_i)\sum_{i}n_i\log(H/\omega_T)+\text{NLf}  \,,
\end{align}
\\
where $n_i=-1,1/2$ for contribution from measure and modes respectively. This is a finite, non- vanishing contribution when $D_1$ has a $1/\delta$ divergence, thus giving a logarithm of total energy. This logarithm is an interesting feature of de-Sitter, which we discuss in greater detail in Sec.~\ref{Sec:Renormalization}.
\\
Consider the $D_1$ contribution to the 1-loop diagram in Fig.~\ref{Figure:general} with quartic ($\dot \pi^4$) and cubic ($\dot \pi^3$) derivative interactions in de-Sitter. Similar to Sec.~\ref{Sec:working_principle_flatspace}, we have loop momenta $p_1$ and $p_2$ flowing between the loop sites $\tau_R$ and $\tau_L$, and each loop site contracts with multiple legs ($p_3,...,p_6$) leading to further tree level sub-diagrams. We refer to the energies flowing through the left blobs into the vertices $\tau_1$ and $\tau_2$ as $\omega_{{L_1}_i}$ and $\omega_{{L_2}_j}$ respectively, and the energies flowing through the right blobs into the vertices $\tau_3$ and $\tau_4$ as $\omega_{{R_1}_k}$ and $\omega_{{R_2}_l}$ respectively.
\\
The partial energies involving the loop momenta are the same as in Eqn.~\eqref{Eqn:partial_general}. 
Evaluation of time integrals in this diagram will lead to all simple poles in partial energies as in Eqn~\eqref{Eqn:loop_gen}, as well as higher order poles due to additional factors of $\tau$ in the numerator coming from modes and measure. The correlator thus gets contributions from terms such as,
\begin{align}
    &\int d^d \vec{p}_1 d^d \vec{p}_2 ~\frac{p_1 p_2 Q(p_i,\omega_j)}{E_1^{\alpha_1} E_2^{\alpha_2} \prod_{ijkl} \left( E_{{L_1}_i}^{\alpha_3} E_{{L_2}_j}^{\alpha_4} E_{L_{ij}}^{\alpha_5} E_{{R_1}_k}^{\alpha_6} E_{{R_2}_l}^{\alpha_7} E_{R_{kl}}^{\alpha_8} \right) } \delta^d (\vec{p}_1 +\vec{p}_2 -\vec{s}) \nonumber \\
    & = \frac{S_{d-2}}{2} \int_{s}^{\infty} dp_+ \int_{-s}^{s} dp_- \frac{p_1 ^{d-2} p_2}{s} \sin^{d-3}{\theta_1}~\frac{p_1 p_2 Q(p_i,\omega_j)}{ E_1^{\alpha_1} E_2^{\alpha_2} \prod_{ijkl} \left( E_{{L_1}_i}^{\alpha_3} E_{{L_2}_j}^{\alpha_4} E_{L_{ij}}^{\alpha_5} E_{{R_1}_k}^{\alpha_6} E_{{R_2}_l}^{\alpha_7} E_{R_{kl}}^{\alpha_8} \right) } \,, \label{Eqn:loop_identity_equality}
\end{align}

where $\alpha_i$'s are integers denoting the order of the poles. The factor of $p_1p_2$ is now in the numerator due to the derivative acting on the modes, i.e. $\dot{f}_k(\tau)\sim \sqrt{k}\tau \exp(i k \tau)$. Note that the limits of the final time integral $\bigg( \int_{-\infty}^0 d\tau \bigg)$ ensures that any additional polynomials in $p_1$ and $p_2$ are not produced, and the only dependence on loop energies are of the form in the previous equation. 

Once again $Q$ is a function of exchange momenta $p_i$ (excluding $p_1$ and $p_2$) and $\omega_j$ (energy of the external legs). Scaling the integrand by $s$ and suppressing the labels $i,j,k,l$, we get the following dependence on $p_-$ and $p_+$,
\begin{align}
    & = s^{m+\delta} 2^{-\delta-1} S_{d-2} \int_{1}^{\infty} d\hat{p}_+ \int_{-1}^{1} d\hat{p}_-~\frac{Q(\hat{p}_i,\hat{\omega}_j) \left(\hat{p}_-^2-\hat{p}_+^2\right){}^2 }{\hat{E}_1^{\alpha_1} \hat{E}_2^{\alpha_2} \hat{E}_{L_1}^{\alpha_3} \hat{E}_{L_2}^{\alpha_4} \hat{E}_L^{\alpha_5} \hat{E}_{R_1}^{\alpha_6} \hat{E}_{R_2}^{\alpha_7} \hat{E}_R^{\alpha_8}} \left(\left(1-\hat{p}_-^2\right) \left(\hat{p}_+^2-1\right)\right){}^{\delta /2} \,,
\end{align}

where $m$ is an integer and the hat denotes scaled momenta.

Comparing the equation above with the flat space result (Eqn.~\eqref{Eqn:loopintegral_flat}) we see that in de-Sitter we have higher order poles (the $\alpha_i$'s), as well as an extra factor of $\left(\hat{p}_-^2-\hat{p}_+^2\right){}^2$, due to $p_1p_2$ being in the numerator as opposed to the denominator. Computing $p_-$ integral, we get,
\begin{align}
\int_{-1}^1 d\hat{p}_- \left(\hat{p}_-^2-\hat{p}_+^2\right){}^2 \left(\left(1-\hat{p}_-^2\right)\right){}^{\delta /2}= \frac{\sqrt{\pi } \Gamma \left(\frac{\delta }{2}+1\right)  \left((\delta +5) \hat{p}_+^2 \left((\delta +3) \hat{p}_+^2-2\right)+3\right)}{4 \Gamma \left(\frac{\delta +7}{2}\right)} \,.
\end{align}

Crucially, the $p_-$ integral above evaluates to polynomials in $p_+$ because of the fact that the loop propagator does not decay faster than $1/p$, where $p$ is the energy running in the loop. The loop propagator will have this behaviour as long as we are working with interactions featuring derivatives on massless scalars. For conformally coupled scalars running in the loop, polynomial interactions are fine as well. However, if a massless scalar runs in the loop leg without a derivative acting on the field, then certain terms in the loop propagator go as $\sim 1/p^3$, and the $p_-$ integral evaluates to hypergeometric functions of $p_+$, which results in the final integral in $p_+$ turning intractable. Also, the equality in Eqn~\eqref{Eqn:loop_identity_equality} can get messed up by complicated interactions, such as inverse laplacians acting on spatial deriatives of the fields (i.e. $\dot{\phi}^2~\partial^{-2}\partial_i\big(\dot{\phi}\partial_i\phi \big)$), which will pull out complicated trigonometric functions. It will be worthwhile to investigate the correlator structure for such interactions, which we leave out for future work.

Using the result of $p_-$ integral and performing partial fraction decomposition, we see that the final integral in $\hat{p}_+$ will be a sum of terms having the following structure,
\begin{align}
    s^{\delta}\int_1^{\infty} d \hat{p}_+ ~ (\hat{p}_+^2-1)^{\delta/2} \frac{\hat{p}_+^y}{\left( \hat{p}_++x \right)^\alpha} 
 \,, \label{Eqn:general_loop_2}
\end{align}

where the $\hat{p}_++x$ in the denominator represents the various partial energies in Eqn.~\ref{Eqn:partial_general} and $\alpha$ is an integer denoting the order of the pole. The values $\alpha$ can take depends on the interacting theory as well as the diagram we are computing, whereas $y$ only depends on the interactions in the theory, e.g. for the derivative interactions we considered in this section $y$ takes values $0,2$ and $4$. Similarly for interactions other than polynomials in flat space, the poles in Eqn,~\eqref{Eqn:general_loop_1} would still be simple, however one would get additional powers of $p_+$ in the numerator, i.e. the loop integral gets the structure of Eqn.~\eqref{Eqn:general_loop_2} with $\alpha=1$. 

To see that the diagrammatic rules apply to various interactions in flat space as well as de-Sitter, notice that Eq.~\eqref{Eqn:general_loop_2} is finite (and hence no logarithms) unless $y\geq \alpha-1$, and for $y,\alpha$ satisfying this inequality with integer $\alpha \geq 1$, Eq.~\eqref{Eqn:general_loop_2} evaluates to the following,
\begin{align}
    = (-1)^{y+\alpha} ~\Mycomb[y]{\alpha-1} ~x^{y\alpha+1} \left[ \frac{1}{\delta} + \log (1+x) +\log s\right] +\text{NLf}  \,.
\end{align}

Once again we see that every partial energy pole turns into a logarithm, with the argument of the logarithm being precisely the partial energy pole at $p_+ \rightarrow s$, exactly as in Eqn.~\eqref{Eqn:bypart_diagrule}. When $x=\hat{E}_1,\hat{E}_2$, the contributions correspond to \textit{loop-subgraphs}, when $x=\hat{E}_{L_1}, \hat{E}_{L_2}, \hat{E}_L$ the contributions correspond to \textit{left-subgraphs}, and finally when $x=\hat{E}_{R_1}, \hat{E}_{R_2}, \hat{E}_R$ the contributions correspond to \textit{right-subgraphs}. 

To conclude this section, we have shown that the rules of Sec.~\ref{Sec:Singularity_structure} work for flat-space as well as de-Sitter. Correlators in de-Sitter also have a contribution $D_2$ coming from $\mathcal{O}(\delta)$ contribution of modes and measure. This produces branch cuts of total energy $\log \omega_T$, which we discuss in greater detail in Sec.~\ref{Sec:Renormalization}.

\subsection{Diagrams with vertices above and below $\tau=0$}
\label{Sec:Unord}
Until now we have been discussing the singularity structure of diagrams with all vertices on the same side of $\tau=0$. However, the correlator will also receive contributions from diagrams in which vertices lie on both sides of $\tau=0$. In this section, we discuss why the rules to extract singularity structure will continue to hold in these diagrams, with an additional constraint on the subgraphs considered.

\begin{figure}[htbp]
    \centering
    \begin{subfigure}{0.44\textwidth}
        \includegraphics[width=\linewidth]{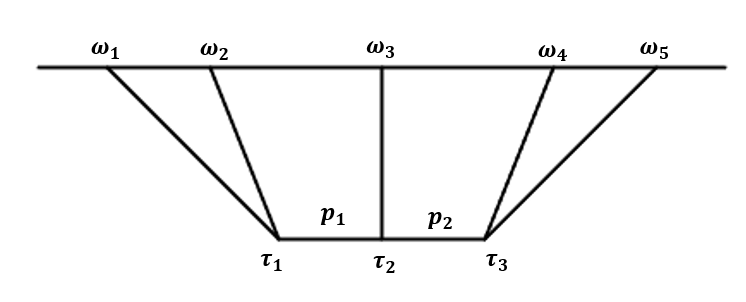}
        \caption{A $5$ pt function at tree level. $\omega_i$ denote external energies, $p_j$ denote exchange legs.}
        \label{Fig:Ord_Tree}
    \end{subfigure}
    \hfill
    \begin{subfigure}{0.54\textwidth}
        \includegraphics[width=\linewidth]{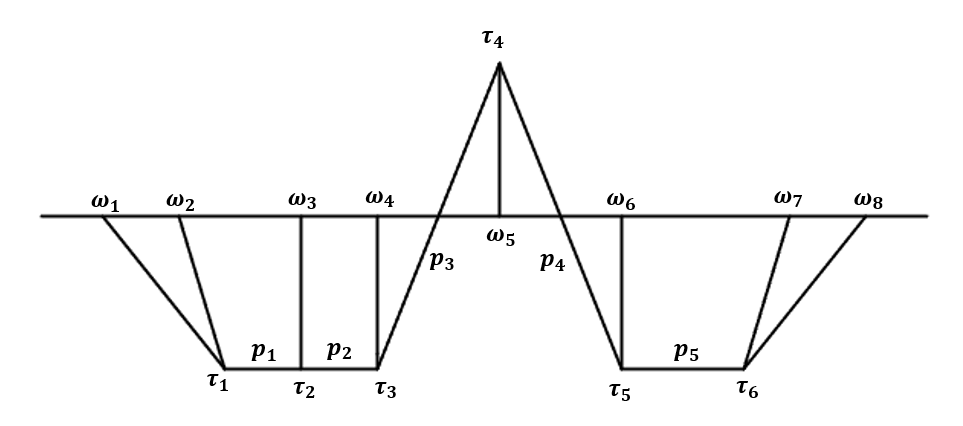}
        \caption{A $8$ pt function at tree level. $\omega_i$ are external legs, $p_j$ denote exchange momenta.}
        \label{Fig:Unord_Tree}
    \end{subfigure}
    
    \caption{Tree level correlator diagrams with a) All vertices below, and b) Vertices both above and below the $\tau=0$ line.}
    \label{Fig:Vertices_Above_Below_Tree}
\end{figure}

In diagrams where all vertices are above or below $\tau=0$, every vertex is connected to its adjacent ones by a bulk-bulk propagator. The Heaviside step function in the propagator causes mixing of energies entering adjacent vertices in the exponential functions, which in turn result in partial energy poles once time integrals are evaluated. To see this, consider the tree diagram in Fig.~\ref{Fig:Ord_Tree}, and let us study the time integral corresponding to one of the time orderings $\Theta(\tau_1-\tau_2) \Theta (\tau_2-\tau_3)$. Considering flat space for simplicity, we have :
\begin{align}
    & \int_{-\infty}^{0}d\tau_1 \int_{-\infty}^{0}d\tau_2 \int_{-\infty}^{0}d\tau_3 ~e^{i (\omega_1+\omega_2-p_1)\tau_1} e^{i (\omega_3+p_1-p_2)\tau_2} e^{i(\omega_4+\omega_5+p_2)\tau_3} \Theta(\tau_1-\tau_2) \Theta(\tau_2-\tau_3) \nonumber \\
    &=  \int_{-\infty}^{0}d\tau_1 \int_{-\infty}^{\tau_1}d\tau_2 \int_{-\infty}^{\tau_2}d\tau_3 ~e^{i (\omega_1+\omega_2-p_1)\tau_1} e^{i (\omega_3+p_1-p_2)\tau_2} e^{i(\omega_4+\omega_5+p_2)\tau_3} \,.
\end{align}

where the Heaviside functions in the first line translate to the changed limits of integration in the second line. From here it is easy to see that the evaluation of the integral on $\tau_3$ pulls a factor of $1/\left( \omega_4+\omega_5+p_2 \right)$ (which is a pole at the energy entering the vertex $\tau_3$), modifying the integrand to $e^{i (\omega_1+\omega_2-p_1)\tau_1} e^{i (\omega_3+p_1+\omega_4+\omega_5)\tau_2}$. Further evaluation of the $\tau_2$ integral pulls a factor of $1/\left(\omega_3+\omega_4+\omega_5+p_1 \right)$ (which is a partial energy pole), modifying the integrand to $e^{i \omega_T \tau_1}$. Finally, the $\tau_1$ integral produces a total energy pole. Clearly, other time orderings will generate other partial energy poles.

However, as pointed out earlier, a $n$-point correlator will also have contributions from diagrams where not all vertices lie on the same side of the line $\tau=0$. Before we study such a diagram at 1-loop diagram, consider the tree level $8$-pt function in Fig.~\ref{Fig:Unord_Tree} in flat space.

Evaluation of time integrals will of course give poles at energies entering the vertices $\tau_1,...,\tau_6$. However, we will not get partial energy poles entering all possible subgraphs. This is simply because $\tau_3,\tau_4$ and $\tau_5$ are connected by Wightman propagators, and hence there are no Heaviside step functions mixing the energies entering these vertices. This implies we will only get partial energy poles for subgraphs with all vertices above \textit{or} below $\tau=0$, i.e. subgraphs where all the vertices are connected by bulk-bulk propagators. 

For example in Fig.~\ref{Fig:Unord_Tree}, for the time ordering $\Theta(\tau_2-\tau_1) \Theta(\tau_2-\tau_3)\Theta(\tau_5-\tau_6)$, we get the following pole structure upon computing time integrals,
    \begin{align*}
       \sim &\frac{1}{\left(\omega_1+\omega_2+p_1\right)}\frac{1}{\left(\omega_4+p_2+p_3\right) } \frac{1}{\left(\omega_5+p_3+p_4\right) } \frac{1}{\left(\omega_7+\omega_8+p_5\right)} \\
       & \times  \frac{1}{ \left(\omega_1+\omega_2+\omega_3+\omega_4+p_3\right) } \frac{1}{\left(\omega_6+\omega_7+\omega_8+p_4\right) } \,,
    \end{align*}
    where the factors in the first line are poles in energies entering the vertices $\tau_1, \tau_3, \tau_4$ and $\tau_6$ respectively, and the factors in the second line are energies entering subgraphs where all vertices are connected by bulk-bulk propagators.

\begin{figure}[htbp]
    \centering
    \begin{subfigure}{0.44\textwidth}
        \includegraphics[width=\linewidth]{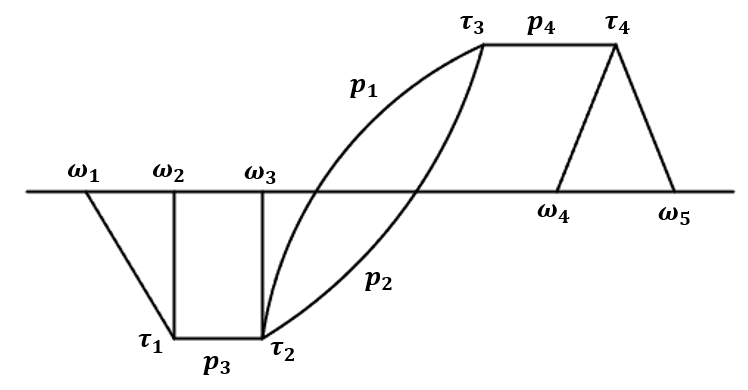}
        \caption{A $5$ pt function at 1-loop. The loop sites are on opposite sides of $\tau=0$.}
        \label{Fig:Unord_Loop_1}
    \end{subfigure}
    \hfill
    \begin{subfigure}{0.54\textwidth}
        \includegraphics[width=\linewidth]{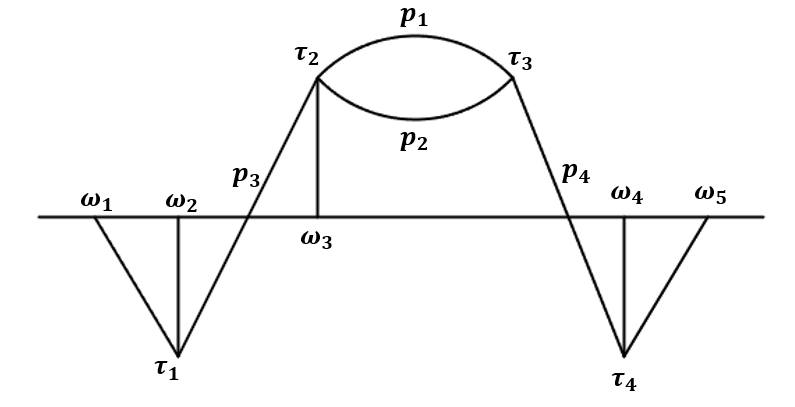}
        \caption{A $5$ pt function at 1-loop. The loop sites lie on the same side of $\tau=0$.}
        \label{Fig:Unord_Loop_2}
    \end{subfigure}
  \caption{Contributions to $5$-point functions at 1-loop with cubic and quartic interactions.}
    \label{Fig:Unord_Loop}
\end{figure}

    Similarly, for 1-loop diagrams with vertices on both sides of $\tau=0$, evaluation of time integrals will produce poles at the energies entering only those subgraphs where all vertices are connected by bulk-bulk propagators. Furthurmore, following the arguments in Sec.~\ref{Sec:working_principle_flatspace} and \ref{Sec:working_principle_deSitter}, the partial energy poles involving loop momentum will turn into branch cuts. Consider Fig.~\ref{Fig:Unord_Loop} as an example. These diagrams, alongwith the one in Fig.~\ref{Fig:Diagram_4}, will contribute to the $5$-pt function at 1-loop with cubic and quartic interactions.

Upon performing time integrals, poles are produced at energies $E_1,E_2,E_3$ and $E_4$, i.e. the energies entering the vertices $\tau_1,\tau_2,\tau_3$ and $\tau_4$ respectively, given in Eqn.~\ref{Eqn:5-pt_1loop}. Evaluation of loop integrals will turn two of these into branch cuts, i.e. $\log E_{2,3}|_{p_+ \rightarrow s}$. Following the rules mentioned in Sec.~\ref{Sec:Singularity_structure}, these are the contributions from \textit{loop-subgraphs}. Time integrals also produce poles at partial energies entering those subgraphs in which vertices lie on the same side, i.e. at $E_L=\omega_1+\omega_2+\omega_3+p_+$ and $E_R=p_++\omega_4+\omega_5$ in Fig.~\ref{Fig:Unord_Loop_1} and at $E_M=p_3+\omega_3+p_4$ in Fig.~\ref{Fig:Unord_Loop_2}. Evaluating the loop integrals will turn the poles at $E_L$ and $E_R$ into branch cuts $\log E_{L,R}|_{p_+\rightarrow s}$, whereas Fig.~\ref{Fig:Unord_Loop_2} does not get any additional branch cuts. The exact same results also follow from the diagrammatic rules, except now while determining the \textit{left-} and \textit{right- subgraphs}, there is an additional constraint that the vertices of the subgraphs should lie on the same side of $\tau=0$. A straightforward but lengthy computation of the diagrams in Fig.~\ref{Fig:Unord_Loop} verifies this.

To summarize, the diagrammatic rules of extracting singularity structure are also applicable to diagrams with vertices above \textit{and} below $\tau=0$, with the additional constraint on \textit{left-} and \textit{right-subgraphs} that all vertices of these subgraphs should lie on the same side of $\tau=0$. Also note that these loop diagram may become finite in de-Sitter if the loop sites are connected by Wightman propagators. If these diagrams are divergent in de-Sitter, then $\log \omega_T$ will also be produced from time integrals, as will be discussed in Sec.~\ref{Sec:Loophole}.

\section{Regularization and analytic structure}
\label{Sec:Renormalization}

Almost everything we have derived so far is in line with expectations from the wavefunction perspective: singularities arise when energy entering a subgraph of the Feynman diagram vanishes. However there is one feature in our example presented in Section~\ref{Sec:de-Sitter examples} which is not accounted for: in Eqn.~\eqref{Eqn:ren_dS_example} the logarithmic terms have total energies in their argument, implying the existence of a total energy branch point. This goes against the prediction of the cosmological KLN theorem \cite{AguiSalcedo:2023nds}, which states: \textit{``There can be no total energy branch points for a cosmological in-in correlator, assuming all interactions are IR finite."}

Here the ``total energy branch point" means the \textit{off-shell} total energy branch point, meaning that given the sum of off-shell energies $\omega_T=\omega_1+\omega_2+\dots+\omega_n$, we cannot have terms like $\log(\omega_T)$ \textit{without} assuming any dispersion relations (such as $\omega_i=|\boldsymbol{k}_i|$). For flat space this is not hard to see: 
\begin{itemize}
    \item Only simple poles can arise from doing time integration. This is a direct consequence of the integrand being a product of $e^{iEt}$.
    \item For a loop diagram, one can write the correlator as a sum of loop level wavefunction coefficients and momentum integrals over tree level wavefunction coefficients. These loop level wavefunctions can be further decomposed into momentum integrals over discontinuities of tree level wavefunction coefficients by using the cosmological Tree theorem.
    \item By adding these discontinuities to the tree level wavefunction coefficients, one finds that it is impossible to obtain denominators of the form $\omega_T+\sum_i a_i p_i$ in the resulting momentum integrand (see Section 3 in \cite{AguiSalcedo:2023nds} for more details). Since logarithmic terms are generated by these denominators by performing the loop integrals, this implies that no total energy logs can appear in the final expression.
\end{itemize}
For general cosmological spacetime with IR finite interactions, the usual expectation is that the correlator (or the wavefunction coefficients) can be found by taking derivatives of the flat space wavefunction with respect to the off-shell energies\cite{Hillman:2021bnk,Lee:2023jby}. Since no total energy branch point can arise from either time or momentum integration, the correlators can only have poles in $\omega_T$, and taking derivatives of poles only give rise to higher order poles. Therefore, the expectation is that the cosmological KLN theorem holds assuming there are no IR divergent interactions.

\subsection{Loophole}
\label{Sec:Loophole}
An assumption in extending the KLN theorem from flat space to general cosmological spacetime is that branch points in energies arise only from momentum integration. However, as we will discuss below, there is a crucial difference between general FLRW spacetimes and flat space. 
\begin{center}\textbf{In de Sitter, total energy branch points in \textit{regularized} correlators can be generated from \textit{time} integrals, rather than momentum integrals.}
\end{center}
For this reason, the arguments of the cosmological KLN theorem are evaded in de Sitter.

\paragraph{Dim Reg}
To see how this happens in dim reg, consider a simple n-point contact diagram for massless scalars in d-dimension dS, with $\lambda'\dot{\phi}^n$ interaction. For $2n-d>0$ this interaction is IR convergent, and usually we write:
\begin{equation}
    \langle \phi^n\rangle=\lambda'\int_{-\infty}^{0}d\tau\, \tau^{2n-d-1}e^{i\omega_T\tau}\sim\frac{\lambda'\Gamma(2n-d)}{\omega_T^{2n-d}}=(-i\partial_{\omega_T})^{2n-d-1}\frac{\lambda'}{\omega_T}.\label{dimreg}
\end{equation} \\
However there is one important caveat: here we must assume the dimension $d$ is an integer, otherwise we are taking fractional number of derivatives. If instead we work in $d+\delta$ dimensions, then we have (substituting the $d$ dimensional mode function):
\begin{equation}
    \langle \phi^n\rangle=(-i\partial_{\omega_T})^{2n-d-1}\lambda'\int_{-\infty}^{0}d\tau\,\tau^{(\frac{n}{2}-1)\delta}e^{i\omega_T\tau}.
\end{equation}

The main difference from flat space is the following: in the expression of Eqn.~\eqref{dimreg} the power of $\omega_T$ is not an integer, therefore $\omega_T=0$ is not a pole (or a higher order pole as one would naively expect), rather it is a branch point (in the same way that $\sqrt{x}$ has a branch point at $x=0$).  Usually we do not concern ourselves with this issue: since $\delta\rightarrow 0$, any corrections are proportional to $\delta$ and it vanishes. However, if there are any divergences, and we work in dim reg, then these correction may give a finite contribution. For example, let $d=3+\delta$, and let $\lambda'=-\frac{\lambda^2}{\delta}$ (i.e. this object is a counterterm to some divergent loop integral), then the integral becomes:
\begin{equation}
    I_{ct}=-\frac{\lambda^2\Gamma(2n-4-\delta)}{\omega_T^{2n-5}}\left(\frac{1}{\delta}+\left(\frac{n}{2}-1\right)\log(\omega_T)+\text{analytic}\right),
\end{equation}
and so we get a $\log(\omega_T)$ contribution. 

At this point it is worth asking what the corresponding divergent loop integral is. In the case where $n$ is an even number, the divergence can arise from a two site loop diagram mediated by the interaction $\lambda\dot{\phi}^{n/2}\sigma^2$, where $\sigma$ is a conformally coupled scalar. The loop integral is quite straightforward to calculate, and it gives:

\begin{align}
    & I_{L}=\lambda^2\partial_{\omega_L}^{n-d-1+2}\partial_{\omega_R}^{n-d-1+2}\left(\frac{\frac{1}{\delta}+\frac{1}{2}\log(\omega_L+s)+\frac{1}{2}\log(\omega_R+s)}{\omega_T^{1+(n/2-2)\delta}}\right)\nonumber\\&=\frac{\lambda^2\Gamma(2n-4-\delta)}{\omega_T^{2n-5}}\left(\frac{1}{\delta}+\left(\frac{n}{2}-2\right)\log(\omega_T)+\frac{1}{2}\log(\omega_L+s)+\frac{1}{2}\log(\omega_R+s)+\text{analytic}\right).
\end{align}
Therefore, the renormalized answer is:
\begin{equation}
    I_{ren}=\frac{\lambda^2\Gamma(2n-4-\delta)}{\omega_T^{2n-5}}\left(\frac{1}{2}\log(\frac{\omega_L+s}{\omega_T})+\frac{1}{2}\log(\frac{\omega_R+s}{\omega_T})\right).\label{renanswer}
\end{equation}

Notice how the argument of the log conspire to form a scale invariant combination: the argument of the log remains unchanged under the scaling $\omega_{L,R,T}\rightarrow \omega_{L,R,T}/a$ and $s\rightarrow s/a$. Similar cancellations also occur in more complicated diagrams, such as the example provided in Section \ref{Sec:de-Sitter examples}. Later we discuss in Section \ref{sec:logcount} that this is a general feature, and for any $n$-point function, the branch cuts always combine to form scale invariant arguments.

The appearance of $\log(\omega_T)$ term is a generic feature of dim reg in dS: the measure for the time integration at each vertex depends on dimension $d$. In addition, since:
\begin{equation}
    \int_{-\infty}^{\tau'}d\tau \tau^{n+\delta}e^{is\tau}=(-is)^{-1-n-\delta}\Gamma(1+n+\delta,-is\tau'), \label{Eqn:time_int}
\end{equation}
by expanding the incomplete gamma function as \cite{NIST:DLMF}:
\begin{equation}
    \Gamma(a,z)=z^{a-1}e^{-z}\sum_{k=0}^{\infty}\frac{(-1)^k(1-a)_k}{z^k}\label{incompletegamma},
\end{equation}
one can see that no partial energy branch points can arise from intermediate time integration (the prefactor $s^{-1-n-\delta}$ in Eqn.~\eqref{Eqn:time_int} cancels the factor of $s^\delta$ coming from the $z^{a-1}$ term in Eqn.~\eqref{incompletegamma}, so partial energies always have integer power). As a result only branch points in $\omega_T$ can arise from time integration, and the power of the singularity $\omega_T$ depends on the number of dimension. Since the number of dimension is no longer an integer in dim reg, naturally this means we get total energy branch points\footnote{This argument relies on the commutation of summation (in Eqn.~\ref{incompletegamma}) and integration operations, however this is tricky. To regularize the loop integrals, $\delta$ must be continued analytically to some point in the complex plane, and it is not immediately obvious that the sum of integrals converges at this value of $\delta$. We will comment on this later.}.

Alternatively, one could argue by expanding the mode functions in $\delta$ before integrating~\cite{Senatore:2009cf}. As an example consider the $D_2$ integral which appears in a two site loop diagram:
\begin{align}
    \int_{-\infty}^{0}d\tau_L\int_{-\infty}^{\tau_L}d\tau_R\,\tau_L^{n_L}\tau^{n_R}_R\log(-H\tau_R)\,e^{i\omega_L\tau_L}e^{i(p_1+p_2)(\tau_R-\tau_L)}e^{i\omega_R\tau_R}.
\end{align}
It was shown in \cite{Bhowmick:2024kld} that for the bispectrum at 1-loop, evaluation of time integrals with the extra factor of $\log \big(-H\tau \big)$ results in a contribution which is equivalent to the logs being pulled outside the integrals as $\log(H/\omega_T)$, plus some rational functions of momenta, as follows,
\begin{align}
    =\log\bigg(\frac{H}{\omega_T}\bigg) \int_{-\infty}^{0}d\tau_L\int_{-\infty}^{\tau_L}d\tau_R\,\tau_L^{n_L}\tau^{n_R}_R \,e^{i\omega_L\tau_L}e^{i(p_1+p_2)(\tau_R-\tau_L)}e^{i\omega_R\tau_R}~+\text{NLf} .
\end{align}

It is straightforward to verify this for higher-point diagrams and conclude that the logarithmic terms generated from expanding the $d-$dimension mode function must have total energy $\omega_T$ as its argument\footnote{Similar to the previous footnote, this relies on an asymptotic expansion of the mode function before doing the integrals, and it is not immediately obvious that the two actions should commute.}. 

\paragraph{Cutoff}
From the argument above, it may seem like these $\log(\omega_T)$ are simply artifacts of working in dim reg, however we can see they will appear in cutoff regularization as well.

Let us again consider the two site bubble example studied previously, and specialize to $n=4$, i.e. we compute a four point correlator for a massless scalar field $\phi$, with the interaction $\dot{\phi}^2\sigma^2$ (where $\sigma$ is a conformally coupled field). The one loop bubble diagram is proportional to the following integral:
\begin{equation}
    I_{L}=\partial_{\omega_L}^2\partial_{\omega_R}^2\int_{-\infty}^{0}d\tau_L\int_{-\infty}^{\tau_L(1+\frac{H}{\Lambda})}
    d\tau_R e^{i\omega_L\tau_L+i\omega_R\tau_R}\int_{s}^{a(\tau_R)\Lambda}dp_+\int_{-s}^{s}dp_- \frac{p_1p_2}{s}\frac{1}{4p_1p_2}e^{i p_+(\tau_L-\tau_R)}.
\end{equation}
Crucially, this is not a derivative of the flat space counterpart of the correlator, due to the additional time dependence of the cutoff. This time dependence is needed, as it is the physical momentum on which the cutoff is imposed. If we evaluate the momentum integral first, we get:
\begin{equation}
    I_{L}=\partial_{\omega_L}^2\partial_{\omega_R}^2\int_{-\infty}^{0}d\tau_L\int_{-\infty}^{\tau_L(1+\frac{H}{\Lambda})}
    d\tau_R e^{i\omega_L\tau_L+i\omega_R\tau_R}\frac{1}{4s(\tau_L-\tau_R)}\left[e^{i\frac{\Lambda}{H}(\frac{\tau_L}{\tau_R}-1)}-e^{is(\tau_L-\tau_R)}\right]\label{cutoffintegral}
\end{equation}
The first term does not give rise to any logarithmic corrections, and for the purposes of studying divergences and logarithmic corrections we may ignore it \cite{Senatore:2009cf,Bhowmick:2024kld}. To convince ourselves that this is indeed the case, we can look at the special case $\omega_R\rightarrow 0$, where the $\tau_R$ integral can be evaluated analytically to give:
\begin{equation}
    \int_{-\infty}^{\tau_L(1+\frac{H}{\Lambda})}
    d\tau_R \frac{1}{4s(\tau_L-\tau_R)}e^{i\frac{\Lambda}{H}(\frac{\tau_L}{\tau_R}-1)}=\frac{1}{4s}\left(e^{\frac{-i\Lambda}{H}}\text{Ei}(\frac{i\Lambda^2}{H^2+H\Lambda})-\text{Ei}(\frac{-i\Lambda}{H+\Lambda})\right).
\end{equation}
Further integration in time will not give rise to any additional logarithmic terms in $\omega_L$, so it cannot possibly generate any total energy logarithms (which in this limit would have the form $\log(\omega_L)$).

The second term integrates to\footnote{The answer from doing the time integration is an exponential integral, which can be expanded in terms of logarithmic terms and a series. By studying the behavior of the exponential integrals at $\tau_L\rightarrow 0$ and $\tau_L\rightarrow -\infty$ one can obtain the answer here.}:
\begin{align}
    \nonumber I_{L}&=\partial_{\omega_L}^2\partial_{\omega_R}^2\int_{-\infty}^{0}d\tau_L\,e^{i\omega_T\tau_L}\text{Ei}(\frac{iH\tau_L(s+\omega_R)}{\Lambda})\\&=\frac{6}{\omega_T^{5}}\left(2\log(\frac{\Lambda}{H})+\log(\frac{\omega_R+s}{\omega_T})+(\text{analytic})\right).
\end{align}
Upon symmetrizing with $\omega_L\leftrightarrow \omega_R$ and renormalizing we obtain the same answer as \eqref{renanswer} (in the special case $n=4$), with the total energy logarithm $\log(\omega_T)$ present.

One may argue that the appearance of logarithmic terms here is an artifact of reversing the order of integration: even if the cutoff does not depend on time, we could still do the momentum integral before the time integral. However, there is a subtle cancellation which occurs when the cutoff does not depend on time:

\begin{align*}
    &\int_{-\infty}^{0}d\tau_L\int_{-\infty}^{\tau_L(1+\frac{H}{\Lambda})}
    d\tau_R e^{i\omega_L\tau_L+i\omega_R\tau_R}\frac{1}{4s(\tau_L-\tau_R)}\left[e^{i\Lambda(\tau_L-\tau_R)}-e^{is(\tau_L-\tau_R)}\right]\\
    &=\int_{-\infty}^{0}d\tau_L\frac{e^{i(\omega_L+\omega_R)\tau_L}}{4s}\left(\text{Ei}(\frac{iH(s+\omega_R)\tau_L}{\Lambda})-\text{Ei}(\frac{iH(\Lambda+\omega_R)\tau_L}{\Lambda})\right)\\
    &=\int_{-\infty}^{0}d\tau_L\frac{e^{i(\omega_L+\omega_R)\tau_L}}{4s}\left(\log(\frac{s+\omega_R}{\Lambda+\omega_R})+\text{analytic}\right).
\end{align*}
Clearly, evaluation of the $\tau_L$ integration will no longer give any total energy logarithms. Hence, we have shown that the total energy logarithm arises from time integrals in de Sitter correlators, irrespective of the regularization method. 

As it is cumbersome to work with cutoff regularisation in higher loops (or more complicated diagrams) we will not try to generalize this result to general loop diagrams.

\subsection{Does the presence of total energy logarithm break scaling}
\label{sec:logcount}
So far we have shown that logarithmic corrections can appear either as $\log \left(  \frac{\omega_p}{\mu}\right)$ or $\log(\frac{\omega_T}{H})$, where $\omega_p$ is some linear functional of energy, and we have restored the factors of $H$ and $\mu$ to ensure that the arguments in the logarithm are dimensionless. In the \textit{on-shell} limit of the correlators, these logarithms are $\log \left(  \frac{k_p}{\mu}\right)$ or $\log(\frac{k_T}{H})$. However, as mentioned in the Introduction, the argument of these logarithmic terms are ratios of comoving and physical scales, and when written in terms of physical scales, they turn into $\log \left( \frac{k}{\mu} \right)=\log \left( \frac{a ~k_{\text{phys}}}{\mu} \right)$. This introduces an explicit dependence on the scale factor for observables, even though the scale factor, by definition, is unobservable.

However, it is not the \textit{unrenormalized} correlators with all its divergences that are relevant to observations. It is the \textit{full, renormalised} correlator that corresponds to physical observables. In the examples we have seen so far (Eqn.~\eqref{renanswer} in Section \ref{Sec:de-Sitter examples}, as well as computations for the Bispectrum presented in \cite{Bhowmick:2024kld}), the arguments of the logarithmic terms are always ratios of physical scales, i.e. 
\begin{equation}
    \log(\frac{k_{p}}{k_{p'}}),
\end{equation}
where $k_p,k_p'$ are linear functionals of energy (for example, it can be $\omega_L+s$, $\omega_R+s$, where $s$ denotes the magnitude of the exchange momentum. It can also be the total energy $k_T$). The log argument is a ratio of two comoving scales, which is perfectly fine from physical arguments\footnote{One way to say this is that the argument of the logarithmic terms are scale invariant, but there are some subtleties about this: see appendix~\ref{appendix_scale_inv_dim_reg} for more details.}. Here we will present a heuristic argument showing why this should always be the case.

First let us note that in a generic loop diagram with massless or conformally coupled scalars, the divergent terms should have this form:
\begin{equation}
    I_{\text{div}}=\sum_i\frac{D^{(i)}_1(k)}{\delta^{l_i}}\left(\frac{k_T}{H}\right)^{-n_i\delta}\prod_{p_i}\left(\frac{k_{p_i}}{\mu}\right)^{m_{p_i}\delta}=\frac{D^{(i)}_1(k)}{\delta^{l_i}}\left(\frac{k_T}{H}\right)^{-(n_i-\sum_{p_i} m_{p_i})\delta}\prod_{p_i}\left(\frac{k_{p_i} H}{k_T\mu}\right)^{m_{p_i}\delta},
\end{equation}
where $k_p$ is some linear functional of external energies labelled by $p_i$, $D_1^{(i)}$ is some scale invariant function in external energies, $l_i$ is an integer which denotes how quickly the term diverges. This form is written with the following consideration:
\begin{itemize}
    \item Since branch cuts at partial energies only arise from momentum integration, factors of $k_p^\delta$ come from the momentum integral measure $\int d^dp $, and $\sum_{p_i}m_{p_i}$ is equal to the dimension of the momentum integrals that produced the branch cuts.
    \item Since branch cuts in total energies are produced only from time integration, thus factors of $k_T^{-\delta}$ come from either the time integration measure ( $\sim \int a^{\delta} d\tau$, hence giving a contribution of $\tau^{-\delta}$) or the mode function (which gives $\tau^{\delta/2}$ thanks to our regularization scheme)\footnote{It is possible to relax this assumption: if we get $k_p^\delta$ instead we can simply rewrite this as $(\frac{k_p}{k_T})^\delta k_T^\delta$, and continue with the analysis. Since $(\frac{k_p}{k_T})^\delta$ is scale invariant this does not affect the result.}. $n_i$ is the sum of these contributions.
\end{itemize}

We would like to cancel the divergences by counterterms, which generically has the following form:
\begin{equation}
    I_{\text{CT}}=\sum_j-\frac{D_1^{(j)}(k)}{\delta^{l_j}}\left(\frac{k_T}{H}\right)^{-(n_j-\sum_{p_j} m_{p_j})\delta}\prod_{p_j}\left(\frac{k_{p_j} H}{k_T\mu}\right)^{m_{p_j}\delta}.
\end{equation}
Note that beyond one loop, we also need to take into account lower order loops as counterterms (for example, at two loops we need to consider one loop diagrams with the one loop counterterms at one of the vertices), which is why generically the counterterms should also depend on partial energies as well.

Properly renormalizing the correlator implies the following:
\begin{equation}
    \sum I_{\text{div}}+I_{\text{CT}}=\text{Finite}.
\end{equation}
Now consider the case where $n_i-\sum_{p_i}m_{p_i}=n_j-\sum_{p_j}m_{p_j}=N$ for all $i,j$. Then this implies, 
\begin{equation}
    \sum I_{\text{div}}+I_{\text{CT}}=\text{Finite and scale invariant}\times \left(\frac{k_T}{H}\right)^{-N\delta},
\end{equation}
since the terms in the remaining brackets are scale invariant. Now expand $\left(\frac{k_T}{H}\right)^{-N\delta}$ as a series in $\delta$. Since it is multiplying a finite and scale invariant function, this implies there can be no $\log\left(\frac{k_T}{H}\right)$ contribution on its own as these are all proportional to $\delta$ and vanishes as $\delta\rightarrow 0$, and so the finite part of the renormalized correlator is scale invariant. It remains to show that $n_i-\sum_{p_i}m_{p_i}=n_j-\sum_{p_j}m_{p_j}$ for all $i,j$. By dimensional analysis, momentum and time integration measure both contribute $k^\delta$, while internal propagators contribute $k^{-\delta}$. By Euler's formula it is straightforward to see that:
\begin{equation}
    n_i-\sum_{p_i}m_{p_i}=\frac{n_{ext}}{2}-L+I-V=\frac{n_{ext}}{2}-1 \,,
\end{equation}
where $n_{\text{ext}},L,I$ and $V$ denote external lines, loops, internal propagators and vertices respectively. This completes the proof.

\begin{figure}[htbp]
    \centering
    \begin{subfigure}{0.44\textwidth}
        \includegraphics[width=\linewidth]{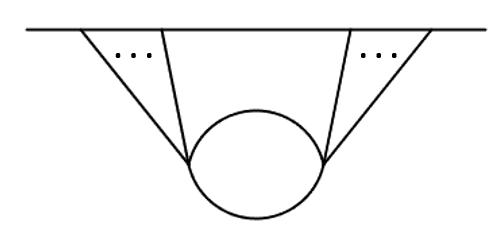}
        \caption{$n$- point diagram at 1-loop with 2-sites. Changing the number of external legs in this diagram would change $n_i$ and $n_j$ in the same way.}
        \label{Fig:NoScaleBreak_1}
    \end{subfigure}
    \hfill
    \begin{subfigure}{0.54\textwidth}
        \includegraphics[width=\linewidth]{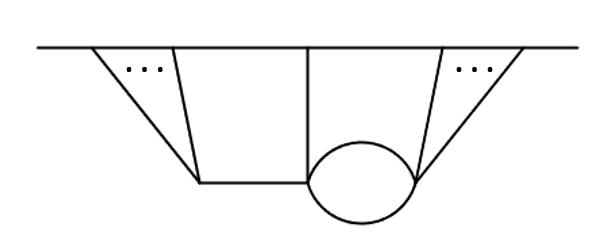}
        \caption{Addition of a new vertex to the diagram in Fig.~\ref{Fig:NoScaleBreak_1} results in an additional time integral which contributes $k_T^\delta$, however this cancels with the contribution coming from the additional propagator.}
        \label{Fig:NoScaleBreak_2}
    \end{subfigure}
    
    \vspace{1em}
    \begin{subfigure}{0.5\textwidth}
        \includegraphics[width=\linewidth]{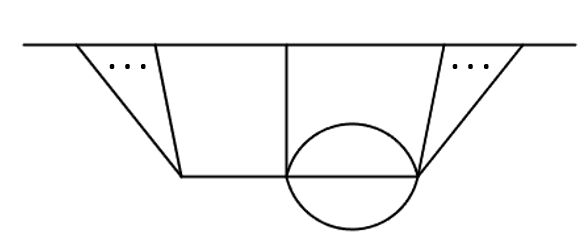}
        \caption{Connecting two sites in Fig.~\ref{Fig:NoScaleBreak_2} increases number of propagators by one, contributing $k_T^{-\delta}$. However this cancels from the contribution arising from the additional momentum integration measure, since this diagram is now a 2-loop diagram.}
        \label{Fig:NoScaleBreak_3}
    \end{subfigure}
    
    \caption{Various $n$ point diagrams at 1- and 2-loops.}
    \label{Fig:NoScaleBreak}
\end{figure}

For clarity let us provide some examples: consider the two site one loop bubble diagram in Fig.~\ref{Fig:NoScaleBreak_1}. Counting the number of $\log(k_T)$, we get
\begin{equation}
    n_i=\frac{1}{2}(n_{ext}+4)-2=\frac{n_{ext}}{2}.
\end{equation}
where $n_{ext}$ denotes the number of external legs. This is because we have 2 internal propagators (each contributing $\tau^{\delta/2\times 2}\propto k_T^{-\delta}$) and two time integrals (each time integral measure contributes $\tau^{-\delta}\propto k_T^{\delta}$ from the measure). 

The momentum integral measure is $\int d^dp$, and by dimensional analysis after doing the momentum integral the measure contributes a factor of $k_s^\delta$, where $k_s$ is the partial energy of one of the vertex. Therefore for the two site one loop bubble diagram, $m_i=1$, and 

\begin{equation}
    n_i-m_i=\frac{n_{ext}}{2}-1
\end{equation}

Now let us renormalise the bubble diagram by adding a counterterm. Here the counter term is just a contact term, which obviously implies $m_i=0$. Since we need to do exactly one time integration, we have:
\begin{equation}
   n_j-m_j=\left(\frac{n_{ext}}{2}-1\right)-0=n_i-m_i,
\end{equation}
and from the discussion above, this result implies the one loop bubble is indeed scale invariant.

Starting from this, it is easy to see why we always have $n_i-\sum_{p_i}m_{p_i}=n_j-\sum_{p_j}m_{p_j}$.
\begin{itemize}
    \item If we add external legs, i.e. we change $n_{ext}$, this changes both $n_i$ and $n_j$ in the same way, so this identity still holds.
     \item Adding a new site in the diagram while remaining at 1-loop does not change $n_i$, since in doing so the new site must be connected with a new propagator (See Fig.~\ref{Fig:NoScaleBreak_2}). While every propagator contributes $k_T^{-\delta}$, every new site introduces a new time integral, and its dimension dependent measure contributes an additional $\tau^{-\delta}\propto k_T^{\delta}$, and so these contributions cancel.
    \item To move to higher loops as in Fig.~\ref{Fig:NoScaleBreak}, connecting two existing sites in a diagram with an additional propagator increases $n_i$ by one (since propagator contributes $k_T^{-\delta}$). However since we have one more loop, there is another loop momentum integration measure $\int d^d p$. By dimensional analysis, after integration this contributes an additional $k_p^{\delta}$, which increases $\sum_{p_i}m_{p_i}$ by one, and so $n_i-\sum_{p_i}m_{p_i}$ remains unchanged. 
    \item The same arguments carry over to the counterterms, so the number $n_j-\sum_{p_j}m_{p_j}$ remains unchanged and is equal to $n_i-\sum_{p_i}m_{p_i}$.
\end{itemize}
So order by order in perturbation, for any arbitrary correlator, scale invariance should be preserved in dim reg.

\section{Conclusion}
In this work, we derive diagrammatic rules to obtain the singularity structure of any $n$-point correlation function at 1-loop with 2-loop sites. These rules bypass the computation of beastly integrals to obtain the poles and branch cuts produced in a diagram, by simple identification of certain subgraphs (\textit{loop-, left-} and \textit{right-subgraphs}) and the energies flowing through them. These rules apply to diagrams with vertices above \textit{and} below the $\tau=0$ line, albeit with an additional constraint on the subgraphs. To demonstrate the robustness of these rules, we show how they follow from the structure of time and momentum integrals of a general 1-loop diagram. 

Additionally, we find that the logarithm of total energy is a feature of \textit{off-shell} correlators in de-Sitter, which we probe in detail. We stress that this branch cut is present in dim reg as well as cut-off regularization. Since this is in conflict with the statement of the cosmological KLN theorem, we revisit the theorem and find a loophole which explains why it does not apply to correlators in de-Sitter. We then show that in a renormalised correlator, the produced branch cuts always repackage themselves into dilatation invariant forms of logarithms at ratios of comoving scales.

Here are a few questions to be addressed in the future.
\begin{itemize}
        \item An obvious criticism to be made is that we have only considered one loop bubble diagrams. The main difficulty here is that the integrands are linear functionals of energies: the integral measure for evaluating this type of functional is significantly more complicated than the ones found in Feynman diagrams (see for instance \cite{Benincasa:2024lxe}\footnote{However progress has been made on the triangle diagram for the wavefunction, see \cite{Benincasa:2024ptf}.}). It would be useful to see how the rules presented in section \ref{Sec:working_principle_deSitter} and \ref{Sec:Loophole} generalise for diagrams at higher loop or larger number of loop-sites. 
    \item It would be interesting to see if correlators with internal fields with arbitrary mass and spins obey the same rules presented in our work. The obvious difficulty here is that obtaining analytic expressions with Hankel function is difficult. One possible way to proceed is to work in Mellin space \cite{Xianyu:2022jwk,Qin:2023bjk,Qin:2024gtr,Sleight:2019hfp,Sleight:2019mgd,Sleight:2020obc,Sleight:2021plv} which makes analytic calculations tractable.
    \item Our computations for the in-in correlators with dim reg relies heavily on expanding the integrand with an asymptotic series before doing the integrals, computing the integral over each term of the series, and then summing it all up. However, it is not immediately obvious that the series expansion commutes with the integrals. Nevertheless, obtaining analytic expressions for correlators under dimensional regularization without resorting to these asymptotic expansions is very difficult, which is the main reason why this technique is often employed in the first place. It would be nice to understand the subtleties of this procedure: for instance, whether we always get $\log(\omega_T)$ corrections as anticipated from our arguments in section \ref{Sec:Loophole}. However it is worthwhile to note that these logarithms arise in cut-off regularisation as well, hence providing a robust cross-check to our results. It is also crucial to investigate the implications of the modifications to the analytic structure introduced by the loop corrections on unitarity and causality, or whether some of these branch cuts are secretly spurious. 
    \item Our rules apply to derivative as well as polynomial interactions in flat space. In de Sitter, if conformally coupled
scalars run in the loop then the rules are valid for both derivative as well as polynomial
interactions. However, if only massless scalars are involved then the rules apply only for derivative interactions, otherwise the loop propagator decays faster than $1/p$, which results in hypergeometric functions upon computing $p_-$ integral. Also, for more complicated interactions, involving for example inverse laplacians acting on spatial derivatives, the general structure of the integrand will be very different from the one considered in this work, and the analysis will be much more complicated. We leave the analysis of correlations with such complicated interactions for a future work.
    \item There exists another dimensional regularization scheme where $d$-dimensional mode functions are used without analytic continuation of the mass (so for instance, the mode function for a massless scalar would be $\tau^{\frac{3+\delta}{2}}H_{3/2+\delta}^{(2)}(k\tau)$). Expanding this mode function in $\delta$ gives additional contributions (see Eqn.~(46) in \cite{Senatore:2009cf} for instance). At one loop, these do not contribute to the logarithmic running  and hence the results from the two regularization schemes agree. It would be nice to check what happens at higher loops, and understand how the heuristic argument presented in \ref{Sec:Loophole} generalises to higher order in perturbation theory.
    \item Finally, it will be interesting to see if these rules imply a simple way to bootstrap the kinematic dependence of one loop correlators of $\zeta$ (curvature perturbation) for full inflationary theory. Using the one loop bubble diagram as an example, we know the type of transcendental functions which can arise from the momentum integration\cite{Salcedo:2022aal,Benincasa:2024ptf}. From our work we understand the location of possible branch points, which in turn fixes the possible arguments of the transcendental functions. It would be interesting to see if we could apply physical constraints such as unitarity~\cite{Goodhew:2020hob,Goodhew:2021oqg,Melville:2021lst,Ghosh:2024aqd,Ghosh:2025pxn} and locality~\cite{Jazayeri:2021fvk} to fix all possible forms of one loop correlators in the same way as tree level wavefunction~\cite{Pajer:2020wxk,Ghosh:2023agt}.
\end{itemize}

\section*{Acknowledgements}
We thank Scott Melville, Santiago Agüí Salcedo, Guilherme Leite Pimentel, Sadra Jazayeri and Elaf Ansari for valuable comments and useful discussions.
DG acknowledges support from the Core Research Grant CRG/2023/001448 of
the Anusandhan National Research Foundation (ANRF) of the Gov. of India. SB acknowledges support through the PMRF Fellowship of the Gov. of India. MHGL is supported a postdoctoral fellowship at the National Taiwan University funded by the Ministry of Education (MOE) NTU-113L4000-1.

\appendix
\section {Scale Invariance}
In flat space one of the main features of dim reg is that it preserves the symmetry of the theory (namely Lorentz invariance and gauge invariance for gauge fields). This is one of the main advantages of using dim reg in flat space over other regularization schemes such as cutoff regularization. However from the examples in \cite{Bhowmick:2024kld} as well as in section \ref{Sec:de-Sitter examples}, we noticed the unrenormalised correlator features logarithms of ratios of comoving momenta and physical renormalization scale $\mu$. Aside from issues with secular divergences mentioned in the introduction, this raises another problem: we expect scale invariance to be a symmetry of the theory, and one may worry if dim reg in de Sitter will violate this. In this Appendix we provide clarification on this issue.\\
``Scale invariance" may refer to a number of things. The first one is the following:
\begin{equation}
    \langle \phi(\lambda x_1,\lambda\tau_1)\phi(\lambda x_2,\lambda\tau_2)\dots\phi(\lambda x_n,\lambda\tau_n)\rangle=\langle \phi(x_1,\tau_1)\phi(x_2,\tau_2)\dots\phi(x_n,\tau_n)\rangle.
\end{equation}
This one always holds as long as the spacetime is well-approximated by de Sitter, and we will not discuss this further.

Cosmologists commonly use scale invariance for a stronger condition: for late time correlators, i.e. $\tau_0\rightarrow 0$, we have
\begin{equation}
    \langle \phi(\lambda x_1)\phi(\lambda x_2)\dots\phi(\lambda x_n)\rangle=\langle \phi(x_1)\phi(x_2)\dots\phi(x_n)\rangle.\label{eqn:scaleinvdef}
\end{equation}
This is a special case of the usual notion of scale invariance in CFT. In cosmology we focus on correlators for massless scalars in $d=3$ at late time, and in this regime massless scalars can usually be decomposed into the following:
\begin{equation}
    \phi(x,\tau_0\rightarrow 0)\sim \mathcal{O}_{\Delta=0}(x)+\tau_0^{3}\mathcal{O}_{\Delta=3}(x).
\end{equation}
Since only $\Delta=0$ survives in the late time limit, demanding scale invariance in the usual CFT notion yields \eqref{eqn:scaleinvdef}.

However, these notions do not always coincide, especially if we work in dim reg. In our version of the regularization scheme we modified the mass such that the order of Hankel function is always a half integer\footnote{If we follow the regularization scheme presented in \cite{Senatore:2009cf} the the CFT notion of scale invariance always coincides with the cosmology one: assuming the mode functions are normalized properly, we always have $\Delta=0$ in their scheme, and \eqref{eqn:scaleinvdef} always holds.}. For massless scalars this implies:

\begin{equation}
    \phi(x,\tau_0\rightarrow 0;\delta)\sim \tau_0^{\delta/2}\mathcal{O}_{\Delta=\frac{\delta}{2}}(x)+\tau_0^{3+\delta/2}\mathcal{O}_{\Delta=3+\frac{\delta}{2}}(x).\label{eqn:phiexpand}
\end{equation}
In our text we are implicitly computing correlators of $\mathcal{O}_{\Delta=\frac{\delta}{2}}$, since we have always stripped off factors of $\tau_0^\delta$ in our computation. This means for our expressions, the correlator scales like the following if we consider the CFT notion of scale invariance:
\begin{equation}
    \langle \phi(\lambda x_1)\phi(\lambda x_2)\dots\phi(\lambda x_n)\rangle=\lambda^{-n\delta/2}\langle \phi(x_1)\phi(x_2)\dots\phi(x_n)\rangle.\label{eqn:scaleinvdefnew}
\end{equation}
The correlators scales as \eqref{eqn:scaleinvdefnew} instead of \eqref{eqn:scaleinvdef}, and this is what people usually mean when they say scale invariance is broken. We refer to the condition \eqref{eqn:scaleinvdefnew} as ``CFT scale invariance".

Fourier transforming \eqref{eqn:scaleinvdefnew} gives,
\begin{equation}
    \langle \phi(\lambda k_1)\phi(\lambda k_2)\dots\phi(\lambda k_n)\rangle'=\lambda^{-d(n-1)+n\delta/2}\langle \phi(k_1)\phi(k_2)\dots\phi(k_n)\rangle',\label{eqn:scaleinvariance}
\end{equation}
where the prime indicates the momentum conserving delta function have been stripped off.

As an example let us consider the bispectrum from a contact diagram mediated by $g\dot{\phi}^3$ interaction, and we take $d=3+\delta$. A straightforward calculation leads to the following:
\begin{align}
    B_3(k_1,k_2,k_3)=\langle\phi(k_1)\phi(k_2)\phi(k_3)\rangle'&=\frac{i^{\delta/2}g\Gamma(3+\frac{\delta}{2})}{8k_1k_2k_3k_T^{3+\delta/2}}\label{eqn:phidot3exact}
\end{align}
Under scaling, we have the following:
\begin{equation}
    B_3(\lambda k_i)=\lambda^{-6-\delta/2}B_3(k_i),
\end{equation}
which indeed scales as \eqref{eqn:scaleinvdefnew} and therefore is CFT scale invariant. Importantly this is true \textit{regardless} of the value of $\delta$ as well as $\lambda$: there is no requirement that $\delta\rightarrow 0$, and $\lambda$ can be as large as we want. 


Should we choose to expand everything as an asymptotic series in $\delta$ things become much harder to keep track of. If we expand \eqref{eqn:phidot3exact} as an asymptotic series in $\delta$, we obtain the following:

\begin{equation}
     B_3(k_i)=\frac{g}{4k_1k_2k_3k_T^3}\left[1-\delta\frac{1}{2}\log(-ik_T)+\dots\right].\label{eqn:ExpandB3}
\end{equation}
Under scaling transformation we have:
\begin{equation}
     B_3(k_i)=\lambda^{-6}B_3(k_i,\delta=0)(1-\frac{1}{2}\delta\log(\lambda)+\mathcal{O}(\delta^2)).\label{eqn:B3scale}
\end{equation}
The additional $\log(\lambda)$ term, despite appearances, cannot be used as indication of CFT scale invariance being broken, and the reason is as follow: expanding \eqref{eqn:scaleinvariance} as an asymptotic series in $\delta$ yields exactly \eqref{eqn:B3scale} at $\mathcal{O}(\delta)$, therefore expression \eqref{eqn:ExpandB3} is still consistent with scale invariance for $d=3+\delta$. Checking full scale invariance requires keeping every single term in the expansion in $\delta$ and properly resumming them, and this is not practical in loop calculations.
\label{appendix_scale_inv_dim_reg}

\bibliography{reference.bib}

\end{document}